\documentclass[fleqn,usenatbib]{mnras}
\usepackage[utf8]{inputenc}
\usepackage{csquotes}
\usepackage{bm}
\usepackage{amsmath}
\usepackage{mathtools} 
\usepackage{amssymb}
\usepackage{graphicx}
\usepackage[usenames,dvipsnames]{color}
\usepackage{hyperref}
\hypersetup{
    colorlinks=true,
    linkcolor=blue,
    filecolor=magenta,      
    urlcolor=cyan,
    pdftitle={Overleaf Example},
    pdfpagemode=FullScreen,
    }
\let\oldhat\hat
\renewcommand{\hat}[1]{\oldhat{\mathbf{#1}}}

\usepackage{bm}
\newcommand*\boldell{\pmb{\ell}}
\newcommand{\non}{\nonumber \\ }
\title[Boosting extragalactic components]{Effects of boosting on extragalactic components: methods and  statistical studies}

\author[William Coulton et al.]{William Coulton$^{1}$, Sydney Feldman$^{2}$, Karime Maamari$^{3}$, Elena Pierpaoli$^{4}$,  \newauthor Siavash Yasini$^{5}$,   and Klaus Dolag $^{6,7}$\\$^{1}$Center for Computational Astrophysics, Flatiron Institute, New York, NY 10010, United States\\$^{2}$Department of Physics and Astronomy,
University of California – Los Angeles, Los Angeles, California 90095, USA\\$^{3}$NASA Langley Research Center, Hampton, Virginia, 23666\\$^{4}$Department of Physics and Astronomy, University of Southern California, Los Angeles, California, 90089-0484 \\$^{5}$Zest AI, Burbank, California,  91505 \\
$^{6}$Universitäts-Sternwarte München, Fakultät für Physik, LMU Munich, Scheinerstra{\ss}e 1, D-81679 München, Germany \\
$^{7}$Max-Planck-Institut für Astrophysik, Karl-Schwarzschild-Str. 1, D-85748 Garching, Germany
}

\date{\today}

\begin{document}

\newcommand{\sy}[1]{\textcolor{blue}{\textbf{\{SY: #1\}}}}
\newcommand{\ep}[1]{\textcolor{red}{\textbf{\{EP: #1\}}}}
\newcommand{\kd}[1]{\textcolor{RubineRed}{\textbf{\{KD: #1\}}}}

\maketitle

\begin{abstract}

In this work we examine the impact of our motion with respect to the CMB rest frame on statistics of CMB maps by examining the one-,  two-, three- and four- point statistics of simulated maps of the CMB and Sunyaev-Zeldovich (SZ) effects.
We validate boosting codes by comparing their outcomes for temperature and polarization power spectra up to $\ell \simeq 6000$.
We derive and validate a new analytical formula for the computation of the boosted power spectrum of a signal  with  a generic frequency dependence. As an example we show how this increases the boosting correction to the power spectrum of CMB intensity measurements by $\sim 30\%$ at 150 GHz. 
We examine the effect of boosting on thermal and kinetic SZ power spectra from semianalytical and hydrodynamical simulations; the boosting correction is generally small for both simulations, except when considering frequencies near the tSZ null. 
For the non-Gaussian statistics, in general we find that boosting has no impact with two exceptions. We find that, whilst the statistics of the CMB convergence field are unaffected, quadratic estimators that are used to measure this field can become biased at the $O(1)\%$ level by boosting effects. We present a simple modification to the standard estimators that removes this bias. Second, bispectrum estimators can receive a systematic bias from the Doppler induced quadrupole when there is anisotropy in the sky - in practice this anisotropy comes from masking and inhomegenous noise. This effect is unobservable and already removed by existing analysis methods.

\end{abstract}

\section{Introduction}

The study of the Cosmic Microwave Background (CMB)  has provided a wealth of information about the composition and evolution of our Universe. Power spectrum measurements of the CMB are the main probe for determining cosmological parameters \citep{collaboration2020planck,Prat_2019, Aylor_2017, choi_act_2020,Bianchini_2020}.  The Sunyaev-Zeldovich (SZ) effects (thermal and kinetic) are also considered in parameter estimation, particularly for $\Omega_m$ and $\sigma_8$ determination \citep{Zubeldia:2019brr}. In the upcoming years, ground--based experiments like advanced ACT \citep{Crowley_2018}, the Simons Observatory \citep{collaboration2019simons}, CMB--S4 \citep{collaboration2020cmbs4} and CMB-HD \citep{sehgal2019cmbhd}  will further these investigations by providing large-area maps in both temperature and polarization with high sensitivity and angular resolution (see Tab. \ref{tab:survey_noise}). The goals of these experiments are very diverse and ambitious, ranging from the detection of primordial B modes at large scales to  the characterization of the growth of the structures through gravitational lensing and  galaxy clusters. Given the exquisite precision of upcoming experiments, the data analysis will present different challenges and opportunities. Extraction of cosmological parameters is expected to happen from the statistical analysis not only of the power spectrum but also of non--Gaussian properties of the various signals. Given the low level of noise, the main challenge in interpreting a given signal will be its separation from the other components. In a regime of such high precision, it is also important to assess the relevance of all possible physical effects and processes that might slightly alter a given signal.

\begin{table}

\begin{tabular}{lccc}
                          & \textbf{$l_{max, T}$} & \textbf{$l_{max, P}$}  & \textbf{sky fraction} \\
                           \hline\hline 
\textbf{advACT}       &  $\ge$  4500  & 4000   & 0.4 \\
\textbf{SO}       & 5000   & 4000   & 0.4 \\
\textbf{CMB-S4}       & 10,000  & 5000   & 0.5 \\
\textbf{CMB-HD }      & 35,000  & - &  0.5      \\
\end{tabular}
\caption{Approximate maximum $l$ modes  measured in temperature and polarization by future experiments, and sky fraction covered \citep{Aiola20,Ade_2019,CMBS4,cmbhd}. }
\label{tab:survey_noise}
\end{table}

The motion of our local frame with respect to the CMB is known to alter our observations because of Doppler and aberration effects. Previous experiments like Planck \citep{collaboration2020planckLVI} were able to measure part of the expected effects, like the prominent temperature dipole \citep{collaboration2020planckLVI, 2014} and mode coupling in the temperature power spectrum \citep{Aghanim:2013suk}. Some other potential implications, such as the alteration of the power spectrum shape and impact on the non--Gaussian signal of the maps, had been deemed irrelevant for the Planck experiment \citep{Catena_2013}. 
The light coming from extragalactic objects such as SZ clusters is also going  to be altered due to our peculiar motion. While some studies have been carried out to address part of the expected observational effects \citep{Chluba2004,Dai_2014,Challinor_2002} a thorough analysis of the impact of such motion on observed
maps  in the microwave and infrared bands is lacking.

The need to take into account the effect of our peculiar motion crucially depends on the experiment's characteristics and overall goals.
For example, The ACT collaboration has in fact corrected its CMB power spectrum for the effect of our motion \citep{Louis_2017}. With new experiments on the horizon, it remains to be addressed whether these corrections are sufficient for a correct data interpretation.
The different area coverage, the noise level (impacting the possibility to measure the CMB on small  scales and in  polarization) and the expanded science goals in terms of astrophysics and cosmology involving galaxy clusters and the use of non-Gaussian signatures  all justify revisiting the issue. Tab.~\ref{tab:survey_noise} shows that in a relatively near future we will be able to map anisotropies out to scales $l \simeq 10,000$. Tab.~\ref{tab:survey_aberration}, reporting a rough measure of the correction needed, suggests that, even for experiments like the Simons Observatory, our peculiar motion cannot be ignored, if subportions of the sky are analyzed separately.

\begin{table}
\begin{tabular}{lc} 
\textbf{Experiment} & $\langle cos\theta \rangle$ \\
                           \hline\hline 
AdvACT & -0.18 \\
SPTPol & -0.44 \\
SPTSZ & -0.25 \\
ACTPol & -0.82 \\
CMB-S4 (Est. whole footprint) & 0.03 \\
LSST & 0.05 \\
SO  (Est. whole footprint)  & -0.03 \\
CMB-HD (Est. whole footprint)  & -0.03 \\
Planck-Gal-70 & 0.01 \\
LOWZ-North & 0.56 \\
DECalS & -0.02 \\
DESI & 0.06 \\
BOSS-DR10 & 0.27 \\
DES & -0.41 \\
BOSS-North & 0.57 \\
CMASS-North & 0.57 \\
eBOSS-North & 0.48 \\
SO  (Est. south footprint) & -0.54 \\
SO  (Est. north footprint)& 0.78 \\
\end{tabular}
\caption{ Measurements performed on patches of the sky that are asymmetric with respect to the boost direction are more significantly impacted by Doppler and aberration effects. We estimate the degree of asymmetry by computing the sky area dependent factor, $\langle \cos \theta \rangle$, for a range of upcoming CMB experiments. Note that the observations for many experiments are actually divided into multiple patches, in order to avoid the galaxy. These separate patches can have significant boost factors as can be seen in the last two rows where we compute the boost factors for the two Simons Observatory (SO) patches.The area coverage for the various experiments cited can be found at: \url{https://github.com/syasini/cmb-x-galaxy-overlaps} }
\label{tab:survey_aberration}
\end{table}

Upcoming experiments plan to exploit the rich cosmological information available in CMB non-Gaussianity (NG) \citep{abazajian2016cmbs4,Ade_2019}. For example studies of NG induced by gravitational lensing aim to provide precision constraints on the amplitude of density perturbations and a detection of the sum of the masses of the neutrino \citep{Allison_2015}. Beyond the cosmological information in lensing itself, lensing induced B modes are a key contaminant for the detection of B modes and need to be accurately removed via delensing \citep{Smith_2012}. Non-Gaussianty from other secondary anisotropies, including the thermal and kinetic SZ effects, is a rich source of cosmological information \citep{Coulton_2018,Crawford_2014}. 

CMB NG is also a potential probe of primordial physics on its own \citep{planck2018_IX,Planck2018VIII,abazajian2016cmbs4}. Current (inflationary) models of the universe assume that the dynamics underlying our observations are Gaussian to a very high degree; however, perfect Gaussianity is impossible with any model \citep{Maldacena_2003,Creminelli_2004}. NG observed in the CMB had to have arisen during the earliest moments of the universe, so measuring NG in the CMB allows us to probe energy scales in the early universe that are unavailable for experimentation otherwise \citep{Chen_2010}. In parallel with the search for primordial gravitational waves, measurements of non-Gaussianity constrain the space of theoretical early universe models \citep{meerburg2019primordial}.  

If boosting CMB signals in the direction of our motion distorts non-Gaussian signals or generates additional noise, it will need to be taken into account when investigating the parameter space of different cosmological models. Conversely, if we show that kinetic effects do not introduce a bias towards non-Gaussianity at these scales, we can confidently attribute measured deviations from Gaussianity to some intrinsic quality of the universe and use this information to motivate future searches, rule out models, and constrain the parameter space.

The goal of this paper is to prepare for the next generation experiments by creating the appropriate tools for boosting/deboosting maps and assessing the expected effects of our motion in the universe on extragalactic components like the CMB and  SZ clusters,  and  properly assessing non-Gaussian statistics of the millimeter sky.

 We first analyze the power spectra of the CMB as well as of the thermal and kinetic Sunyaev Zel’Dovich effects (tSZ and kSZ) in the rest and boosted frames \citep{Sunyaev_1972,Sunyaev_1980}. The effect of our motion on the CMB has been investigated using three different methods: real space boosting \citep{Yoho:2012am}, Fourier space boosting \citep{Dai_2014}, and a boosting approximation applied directly to the power spectrum \citep{Jeong:2013sxy}. These methods are explained in further detail in Section \ref{sec:boostingMethods}. However, these analyses have only been performed up to, at most, $\ell$ = 3000 for the CMB, and not at all for the frequency dependent observables, such as the SZ. We are motivated by upcoming high resolution surveys, such as those given in Tab. \ref{tab:survey_noise}, and theoretical advances on the frequency-dependent boosting \citep{Yasini:2017jqg} to extend our investigation to $\ell$ = 6000 (including polarization) and to the SZ effects, with the goal of assessing expected  signatures  of our motion at high resolution and a broader range of observables. 

In order to meet these objectives, we generalize the expression of the power spectrum boosting formula, originally developed for rest-frame black body emission, to any kind of emission law. 
By doing so, we can apply the boost to maps of tSZ effect, which has its own frequency dependence \citep{Sunyaev_1980}. 
We then assess the effects of the boosting using actual sky simulations of the CMB and the SZ effects. This allows us 
to verify the accuracy of the three boosting methods (real-space, Fourier-space, and power spectrum boost), specifically at high ells, and compare their respective computational efficiencies for future use.

Second, we use non-Gaussian simulations to consider how higher order statistics are impacted by the Doppler and aberration effects. We analyze a range of one point statistics of the spherical harmonic coefficients from CMB, tSZ and kSZ maps to determine if our motion induces or modifies non-Gaussianity in these data. Next we consider how measurements of the primordial bispectrum and of the tSZ and kSZ bispectra are impacted by boosting effects. Finally we consider whether inferences of the CMB lensing potential, a four-point function of the CMB maps, is impacted.

The paper is structured as follows: In Section \ref{sec:physicalEffects} we provide a mathematical overview of the relevant physical effects: the Doppler and aberration effects, as well as integrated Sachs Wolfe and gravitational lensing.  In Section \ref{sec:boostingMethods} we describe the three boosting methods we compared, including the advantages and limitations of each method. In Section \ref{sec:simulations} we provide details of the CMB simulations and the two SZ simulations used in this work. We next validate these methods using high--resolution CMB maps, Section \ref{sec:twoPoint}. In Section \ref{sec:onePointAn}, we use a range of one-point statistics to study the impact of Doppler boosting on the CMB and the SZ effects. In Section \ref{sec:threePointAn} and Section \ref{sec:fourPointAn} we investigate how bispectrum measurements and lensing, four-point function analyses are impacted by the aberration effects. We conclude in Section \ref{sec:conclusions}.

\section{Physical effects: an overview}\label{sec:physicalEffects}
In this Section we briefly review the main physical effects discussed in this work: the Doppler and aberration effects, Sec. \ref{sec:dopplerAndAber}, the Sunyaev Zel'dovich effects, Sec \ref{sec:SZeffects}, and the integrate Sachs-Wolfe and CMB lensing effects, Sec. \ref{sec:CMBeffects}. 
\subsection{Doppler and aberration effects}\label{sec:dopplerAndAber}
The two ways in which observations are influenced by our particular velocity are the Doppler and the aberration effects. In an inertial frame moving at relativistic speeds, the Doppler effect causes the frequency of light to appear shifted in the following way:
\begin{equation} \label{eq:DopplerEffect}
    \nu^{\prime} = \gamma \nu (1 + \beta \cos\theta)
\end{equation}
where $\beta$ is the velocity of the observer’s frame of reference in speed of light units, $\gamma =1/ \sqrt{1-\beta^2}$ is the Lorentz factor, $\nu$ is the frequency in the rest frame of the source, $\nu^{\prime}$ is the wavelength in the observer’s frame, and $\theta$ is the angle between the velocity of the observer and the incident angle of the light in the rest frame of the source. Throughout this work we noted quantities in the boosted frame with primes and tildes e.g. $\tilde{T}(\mathbf{n}')$ for the CMB temperature in the boosted frame, and quantities in the rest frame without.

The aberration effect changes the observed direction at which the light is detected. The transformation between the direction of the light in the rest frame of the source and the observed direction of the light in the moving frame can be described by the following equation:
\begin{equation} \label{eq:aberationLocations}
   \hat{n^{\prime}} = \frac{\cos\theta + \beta}{1 + \beta \cos\theta}\hat{\bm\beta} + \frac{\hat{n}-\hat{\bm\beta} \cos \theta}{\gamma(1 + \beta \cos \theta)}
\end{equation}
where $n$ is the unit vector pointing in the direction of the light in the rest frame of the source and $n^{\prime}$ is the unit vector pointing in the direction of the observed light.

\subsection{Sunyaev-Zeldovich effects}\label{sec:SZeffects}

The 
SZ effects occur when CMB photons 
interact with hot electrons in the late-time Universe, like the ones in galaxy clusters.
There are two mechanisms in which CMB photons are affected by these interactions. The first mechanism is through inverse Compton scattering, where photons absorb thermal energy from high temperature electrons as they pass through the interstellar medium. This effect, known as the thermal Sunyaev-Zel’Dovich effect (tSZ) \citep{Sunyaev_1972,Sunyaev_1980}, produces a spectral distortion of the CMB spectrum given by 
\begin{equation}
    \frac{\Delta T^{\rm{tSZ}}(\hat{n},\nu)}{T_{\rm{CMB}}} = y \operatorname{g}\left(x\right)
\end{equation}
where $x=\frac{h \nu}{k_{B} T_{\rm{CMB}}}$, $\nu$ is the observation frequency , $h$ is Planck's constant,  $k_B$ is the Boltzmann constant,  $T_{\rm{CMB}}$ is the CMB temperature, $y$ is the (dimensionless) Compton $y$-parameter and $g$ is the tSZ response function. The tSZ response function characterizes the frequency dependence of spectral distortion as
\begin{equation}
    \operatorname{g}\left(x\right)={x}\,{\coth\left(\frac{x}{2}\right)}-4.
\end{equation}
The Compton $y$-parameter depends on the cluster's characteristics through  
\begin{equation} 
    y = \int{\mathrm{d}\chi a(\chi) \sigma_T \frac{P_e (\chi \unboldmath{\hat{n}})}{m_e c^2}}
    \label{eq:yparam}
\end{equation}
where $\sigma_T$ is the Thomson cross section, $a$ is the scale factor, c is the speed of light, $\chi$ is the comoving distance,  and $m_e$ and $P_e$ are the electron mass and electron pressure.

In the second mechanism, electrons transfer kinetic energy from their bulk motion to incoming CMB photons \citep{Sunyaev_1972,Rephaeli_1991}. This effect, known as the kinetic Sunyaev-Zel’Dovich effect (kSZ), is an order of magnitude smaller than the tSZ and produces spatial anisotropies with the same frequency spectrum as the primary CMB - this makes it more difficult to disentangle these anisotropies from the primary CMB anisotropies. Relative to the CMB, temperature fluctuations caused by the kSZ are given by
\begin{equation} \label{eq:kSZeffect}
    \frac{\Delta T^{\rm{kSZ}}(\hat{n})}{T_{\rm{CMB}}} = - \int{\mathrm{d}\chi a(\chi) \sigma_T n_e(\chi) \mathbf{v} (\chi \mathbf{\hat{n}}) \cdot \mathbf{\hat{n}}  e^{-\tau(\chi)}}
\end{equation}
where $\bf{v} $ is the electron velocity and $\tau$ is the optical depth. 

Data from the tSZ and kSZ have been proposed as cosmological parameter probes themselves, and are used to accurately detect galaxies. In particular tSZ cluster counts  particular have been investigate the tension between the CMB and large scale structures  of the universe \citep{Leauthaud_2017,Zubeldia:2019brr} and measurements of the kSZ effect are powerful probes of astrophysics and primordial non-Gaussianity \citep{Schaan_2021,Munchmeyer_2019}. The SZ effects have proven extremely significant to the study of cosmology and will only become more relevant with higher resolution surveys. Therefore, it is important to understand how our motion in the universe may impact our observations of these effects, especially at small angular scales.

\subsection{CMB effects - the integrated Sachs Wolfe effect and gravitational lensing}\label{sec:CMBeffects}
Gravitational lensing perturbs the trajectories of CMB photons as they propagate from the last scattering surface (LSS) to the observer \citep{Blanchard1987,Bernardeau1997}. These deflections mean that the CMB anisotropies at LSS, $\Delta\bar{T}(\mathbf{n})$, are remapped so that they are related to the observed temperature anisotropies, $\Delta{T}(\mathbf{n})$, by \begin{equation}\label{eq:lensingMapLevel}
 \Delta{T}(\mathbf{n}) = \Delta\bar{T}(\mathbf{n}+\nabla \phi(\mathbf{n})).
\end{equation}
$ \phi(\mathbf{n})$ is the lensing potential and is defined as
 \begin{equation}\label{eq:lensingPoten}
  \phi(\mathbf{n}) = -2 \int_0^{\chi_{*}}\mathrm{d}\chi\frac{\chi_*-\chi}{\chi_*\chi} \Phi(\chi\mathbf{n},\chi).
\end{equation}
where $\Phi$ is the gravitational potential and $\chi_*$ is the comoving distance to LSS. This remapping introduces non-Gaussianity to the CMB anisotropies; specifically a non-vanishing four-point function. This effect does not distort the CMB spectrum. The size of the CMB deflections depends upon the integral of the gravitational potential, and hence the density fluctuations,  from LSS to the observer, weighted by a geometric factor. This geometric factor means that the deflections are most sensitive to perturbations at redshifts between $z=0.5$ and $z=3.$ \citep{Zaldarriaga_1999,LEWIS_2006}. We discuss how lensing can be measured, and how it is impacted by Doppler and aberration effects in Section \ref{sec:lensing4point}.

In additional to deflection, CMB photons can also be redshfited by gravitational potentials as they propagate through the Universe. This effect, known as the integrated Sachs-Wolfe (ISW) effect on linear scales and the Rees-Sciama effect on non-linear scales, generates temperature anisotropies whose amplitude is given by
\citep{Sachs_1967,Rees_1968,Martinez-Gonzalez_1990}
\begin{equation}
    \frac{\Delta T(\mathbf{n})}{T_{\rm{CMB}}} =  {-2}\int^{\chi_{*}}_0 \mathrm{d}\chi \frac{\partial \Phi(\chi\mathbf{n},\chi)}{\partial\chi}.
\end{equation}
The growth of potentials under gravitational collapse, and the decay of potentials through dark energy driven expansion, thus leaves an imprint on the CMB and is a significant contribution to the large scale CMB temperature power spectrum. The correlation between the ISW effect and the lensing effect - as both depend on the line-of-sight gravitational potential - introduces further non-Gaussianity into the CMB, which is explored further in Section \ref{sec:ISWlensingBispectrum}.

\section{Boosting Methods}\label{sec:boostingMethods}

When dealing with our peculiar motion, we may be faced with the task of estimating the boosted power spectrum on a given area of the sky, or producing the boosted image, given the restframe one, and then computing all relevant statistics.
The first problem has been addressed, in the simplified case of a black body spectrum, in \citet{chluba2011aberrating,Jeong:2013sxy}; and it is revisited here in Sec.~\ref{sec:Jeong}.

The second problem has also been discussed in the literature. In this Section we review the proposed boosting methods and then we perform a thorough analysis 
of the methods' performances in Sec.~\ref{sec:twoPoint}.

\subsection{Boosting at the power spectrum} \label{sec:Jeong}

The most computationally efficient way to obtain  the boosted CMB power spectrum ($C_\ell^\prime$) from the rest frame one  is to apply the following analytical formula derived  in \citet{Jeong:2013sxy}:
\begin{equation}
C_\ell^\prime = C_\ell \left(1+ \frac{\Delta C_\ell}{C_\ell}\right)
\label{eq:Clboosted}
\end{equation}
Where
\begin{equation}
\frac{\Delta C_\ell}{C_\ell}=-\frac{d \ln C_\ell}{d \ln\ell}\beta \langle \cos\theta \rangle + O(\beta^2)
\label{eq:Jeong}
\end{equation}
where the angle average is taken over the area covered by the survey.
This is an approximation of the modulation and aberration effects to first order in $\beta$, where $\beta$ is our velocity relative to the CMB and is equal to 0.00123 (in speed of light units) \citep{Aghanim:2013suk}. 
This method has been shown to be accurate when applied to CMB temperature maps up to $\ell = 3000$ through extensive simulations \citep{Jeong:2013sxy}.
It also takes into account the masking function of a particular map in a computationally efficient way.
An evaluation of the average $\cos(\theta)$ for current and future surveys is provided in Tab.~\ref{tab:survey_aberration}.
While the effect of our motion is greatly reduced when large areas of the sky, in symmetric directions with respect to our motion are surveyed, the effect can be quite substantial if sub-areas are considered independently.

This formula is only applicable to frequency independent signals and those with Doppler-weight one,
where an observable, $F(\nu,\mathbf{n}$), with Doppler-weight, d, transforms under a boost as
\begin{align}
    \frac{F'(\nu',\mathbf{n}')}{\nu'^d} =  \frac{F(\nu,\mathbf{n})}{\nu^d}.
\end{align}

We extended the power spectrum boosting formula to account for frequency dependent observables and those with Doppler-weight $d\ne1$. This is important for two reasons: firstly it is necessary to study how our motion impacts the power spectrum of CMB secondary anisotropies. Secondly, 
all current CMB experiments do not directly measure the temperature anisotropies and instead they measure intensity fluctuations of the sky, $\Delta I(\nu,\mathbf{n})$, a frequency dependent quantity. The CMB has a black-body spectrum and, using the fact the CMB anisotropies are small perturbations to the CMB temperature, the measured intensity fluctuations can be linearly related to the temperature anisotropies
\begin{align}
    \Delta I(\nu,\mathbf{n}) = \frac{2 h^2 \nu^4}{c^2k_B T_{\mathrm{CMB}} ^2}\frac{exp\left[\frac{h\nu}{k_B T_{\mathrm{CMB}} } \right]}{\left( exp\left[\frac{h\nu}{k_B T_{\mathrm{CMB} }} \right]-1 \right)^2} \Delta T(\mathbf{n}).
\end{align}
Thus to compute the impact of our motion on the measured CMB anisotropies we need to understand how the power spectrum of frequency dependent obserables is impacted.
 Here after CMB measurements performed this way are referred to as `differential thermodynamic measurements' and we refer the reader to \citet{Aghanim:2013suk,collaboration2020planckLVI} for more details.

To extend the formula, first consider the case of a frequency independent, Doppler-weight 1 (d=1) field. The power spectrum in the boosted frame is given as
\begin{align}
    C'_{Lm} & = \langle {a'}_{LM}^*a'_{LM} \rangle \nonumber \\
    &=  \sum_{\ell,\ell'} \langle {}^1 \mathcal{K}_{L\ell}^{m*} a^*_{\ell m} {}^1\mathcal{K}_{L\ell'}^{m}a_{\ell' m} \rangle
\end{align}
where for simplicity we aligned the boost with the $\hat{z}$ direction and ${}^1 \mathcal{K}_{L\ell}^{m*}{}^1$ are the boosting kernels introduced in \citet{Challinor_2002}. These can be written as
\begin{align} \label{eq:kernelDef}
    {}^1\mathcal{K}_{L\ell}^{m} = \langle Lm| W e^{i\eta \hat{Y}_z }| \ell m \rangle.
\end{align}
using the operator notation introduced in \citet{Dai_2014} where $\eta$ is the rapidity, $\beta = \tanh \eta$, , $W$ is the sky mask and $\hat{Y}_z$ is the boosting operator. 
Expanding to leading order we have
\begin{align} \label{eq:boost_d1_op}
    C'_{Lm} & \approx C_{Lm} +\sum_{\ell} \left[ \langle LM|W|\ell m\rangle \langle Lm| W {i\eta \hat{Y}_z }| \ell' m \rangle^* \right. \nonumber \\ & \left.+ \langle LM|W|\ell m\rangle^* \langle Lm| W {i\eta \hat{Y}_z }| \ell m \rangle  \right]C_{\ell m}.
\end{align}
\citet{Dai_2014} found that this is accurately approximated by
\begin{align}
    C'_{Lm} & \approx C_{Lm}-\beta \langle \cos\theta \rangle \frac{\mathrm{d}C_{Lm}}{\mathrm{d}\ln\ell}.
\end{align}
For  frequency-dependent, general Doppler-weight fields we have
\begin{align}
    C'_{Lm}(\nu_1,\nu_2) & = \langle {a'}_{LM}^*(\nu_1'){a'}_{LM}(\nu_2') \rangle \nonumber \\
    &=  \sum_{\ell,\ell'} \langle {}^d \mathcal{K}_{L\ell}^{m*}(\nu_1') a^*_{\ell m}(\nu_1') {}^d\mathcal{K}_{L\ell'}^{m}(\nu_2')a_{\ell' m}(\nu_2') \rangle
\end{align}
As was shown in Chluba et al (in prep) the new kernels are given by
\begin{equation}
     {}^d \mathcal{K}_{L\ell}^{m*}(\nu_1)=\langle Lm| W  e^{i\eta\hat{Y}^d_z} | \ell m \rangle
\end{equation}
where the new boost operator is given by
\begin{equation}
    \hat{Y}^d_z  = \hat{Y}_z - i\mu (d-1 -\nu \partial_\mu)
\end{equation}
Replacing the boost operator in Eq. \ref{eq:boost_d1_op} with the more general boost operator straightforwardly gives
\begin{align}\label{eq:jeongExtended}
 &    C'_{Lm}(\nu_1',\nu_2') \approx C_{Lm}(\nu_1',\nu_2')-\beta \langle \cos\theta \rangle C_{Lm}(\nu_1',\nu_2')\times  \nonumber \\ & \left[ \frac{\mathrm{d}\ln C_{Lm}(\nu_1',\nu_2')}{\mathrm{d}\ln\ell} +2(d-1) -\frac{\mathrm{d}\ln F(\nu_1')}{\mathrm{d}\ln \nu_1'} -\frac{\ln F(\nu_2'))}{\mathrm{d}\ln \nu_2'} \right],
\end{align}
having assumed that the frequency and spatial dependencies are separable so the total (cross-) spectrum can be written as 
\begin{equation}
C_{\ell m} = F(\nu_1)F(\nu_2) G(\ell,m),
\end{equation}
where $F(\nu)$ and $G(\ell,m)$ are arbitrary functions. 

Second order aberration boost terms are generally negligible, as the $ \ell$ derivatives of the power spectrum are generally much less than $1/\beta$ - this is explicitly seen in Section \ref{sec:twoPoint} where we see agreement between the first order boosting formulae and the numerical methods for $\ell<5000$. However higher-order frequency derivatives can become important, for example when the frequency response has a null and deep in the Wien tail. Including only the dominant second order terms (those second order in frequency derivatives) we have
\begin{align} \label{eq:jeongExtendedWith2ndFreq}
     &C'_{Lm} \approx C_{Lm}-\beta \langle \cos\theta \rangle C_{Lm}\left[ \frac{\mathrm{d}\ln C_{Lm}}{\mathrm{d}\ln\ell}+2(d-1)-\frac{\mathrm{d}\ln F(\nu_1)}{\mathrm{d}\ln \nu_1} \right. \nonumber \\ 
    & \left. -\frac{\ln F(\nu_2)}{\mathrm{d}\ln \nu_2} \right]
    +\beta^2C_{Lm}\langle \cos\theta\rangle^2 \frac{\mathrm{d}\ln F(\nu_1)}{\mathrm{d}\ln \nu_1}\frac{\mathrm{d}\ln F(\nu_2)}{\mathrm{d}\ln \nu_2} \nonumber \\ & +\frac{1}{2}\beta^2C_{Lm}\langle \cos^2\theta\rangle \left[ \frac{\mathrm{d}^2\ln F(\nu_1)}{\mathrm{d}\ln \nu_1^2}+\frac{\mathrm{d}^2\ln F(\nu_2)}{\mathrm{d}\ln \nu_2^2} \right].
\end{align}
Note that the first order expansion, Eq. \ref{eq:jeongExtended}, is sufficiently accurate for most purposes and thus it will be our baseline power spectrum boosting method. In Section \ref{sec:twoPoint} we examine how well these updated formulae work on maps the of tSZ effect and differential thermodynamic measurements of the CMB. 

\subsection{Boosting at the map level} \label{sec:boostmet}
We will now discuss boosting at the map level, which we perform via two approaches: in real space and harmonic space.

\subsubsection{\textit{Pixell} - a real-space boosting code}
The first boosting method involves applying the Lorentz transform to CMB maps in real space. The boost is applied at the pixel level, mapping each angular and frequency data point in the rest frame to a new angle and frequency in the boosted frame. 

Our real space method uses the publicly available \textit{Pixell} library \footnote{\url{https://github.com/simonsobs/pixell}}. The boosted temperature anisotropies are generated by directly evaluating
\begin{align} \label{eq:boostedDeltaT}
\tilde{T}(\mathbf{n}') = \frac{ {T}(\mathbf{n})}{\gamma(1-\beta \cos \theta')},
\end{align}
where again tildes and prime denote quantities evaluated in the boosted frame.

Specifically, we first generate a Gaussian unboosted CMB, $\Delta T(\mathbf{n}_i)$ where $\mathbf{n}_i$ is the set of pixels. The code then uses this pixel locations as the pixel locations of the output, boosted map. To evaluate the aberration we first compute the location of these output pixels in the unboosted frame; 
these locations are given by evaluating Eq. \ref{eq:aberationLocations} with -$\beta$. We then use bi-cubic interpolation to evaluate the simulated unboosted CMB at these new positions. The Doppler term can then be computed by multiplying the aberrated map by $1/(\gamma(1-\beta \cos\theta'))$, which is trivial to evaluate.

The \textit{Pixell}  library can also compute boosted CMB anisotropies for differential intensity measurements. For differential intensity measurement at frequency, $\nu'$, the boosted maps are evaluated by directly computing 
\begin{align} 
   &  \Delta\tilde{T}(\mathbf{n}',\nu') = \frac{\Delta {T}(\mathbf{n})}{\gamma(1-\beta \cos\theta')} 
    +\left[\left( \frac{1}{\gamma(1-\beta \cos\theta')}-1\right)^2 T_{{\rm CMB}} \right.  \nonumber \\ & 
    \left. + \frac{2}{\gamma(1-\beta \cos\theta')}\left( \frac{1}{\gamma(1-\beta \cos\theta')}-1\right)\Delta {T}(\mathbf{n})  \right]\left(g(x')+3\right) \nonumber \\ & 
    +\left( \frac{1}{\gamma(1-\beta \cos\theta')}-1\right)^2 T_{{\rm CMB}}  + \frac{g(x')+3}{[\gamma(1-\beta \cos\theta')]^2T_{{\rm CMB}} }\Delta {T}(\mathbf{n})^2.
\end{align}
The four terms correspond to the standard modulation, the quadrupole term, the dipole term and the second order anisotropy term. The later term is a negligible term.

This method requires high resolution maps in order to have accurate interpolations. With low resolution maps the effect of the pixel window function, which suppresses power when the interpolated data points do not map directly to the middle of a pixel in the boosted frame - see \citep{Yoho:2012am} for a more detailed discussion.  Note that the \textit{Pixell} interpolation method has a significantly reduced transfer function compared to the method implemented in \citet{Yoho:2012am}.

\begin{figure*}
    \centering
    \includegraphics[width=0.49\linewidth]{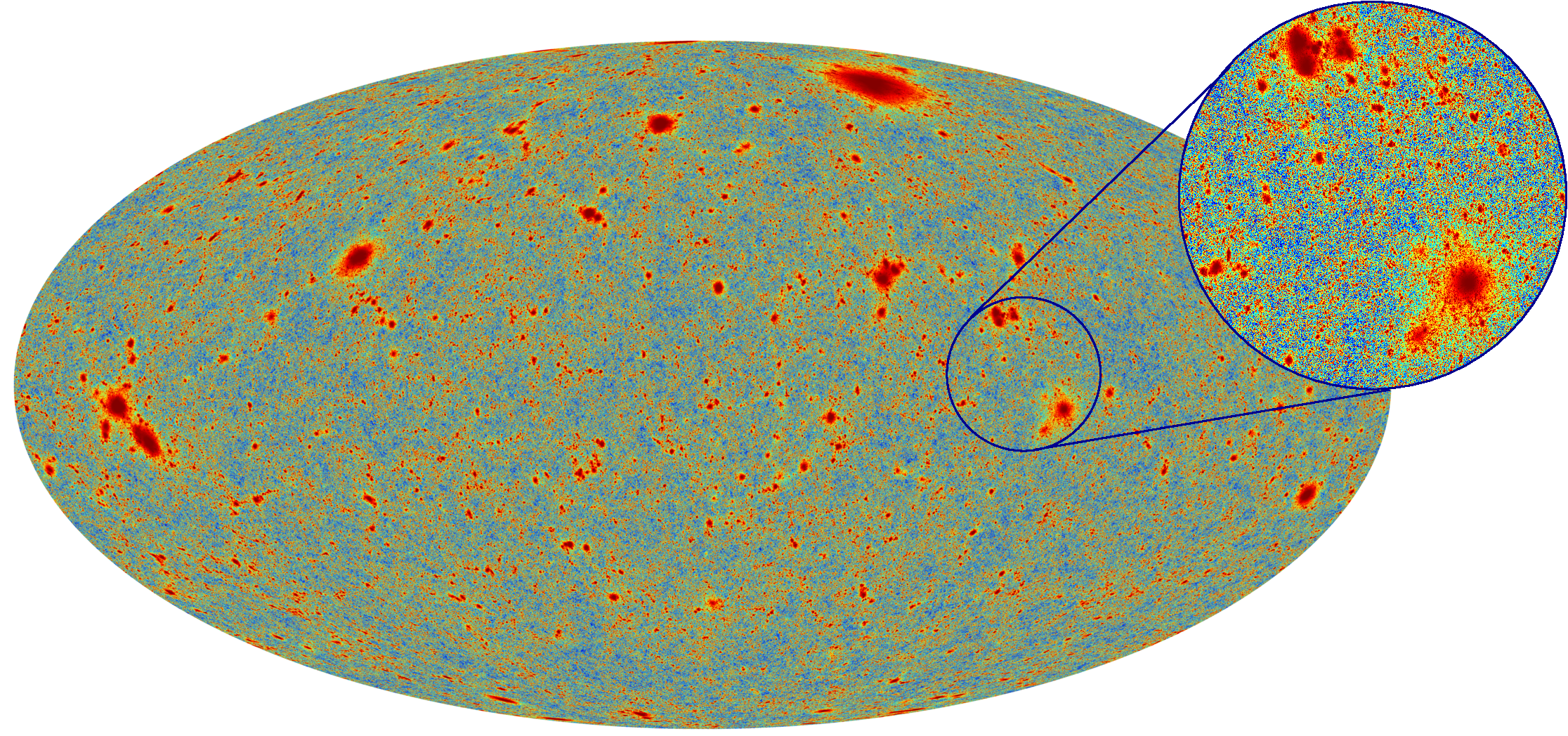}
    \includegraphics[width=0.49\linewidth]{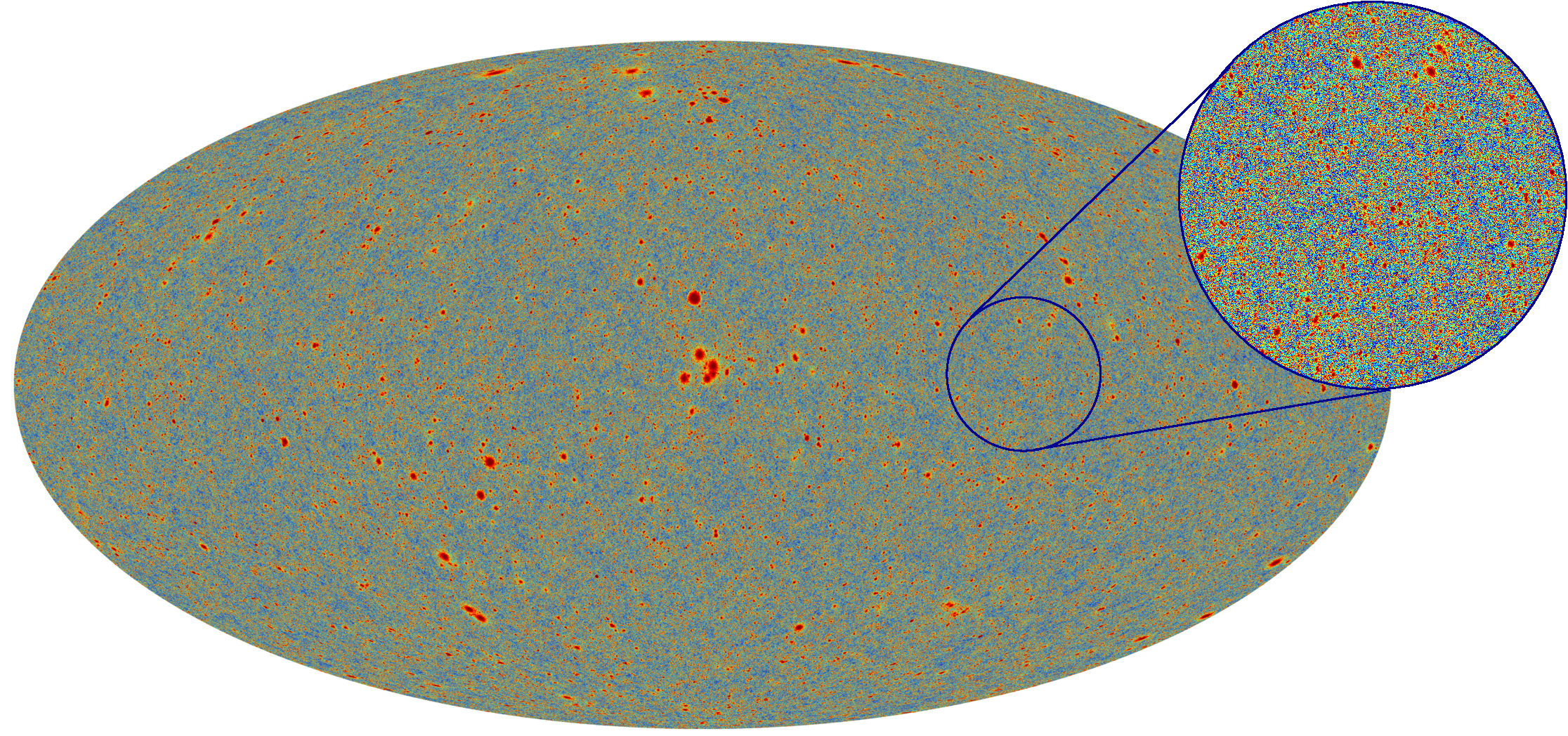}\\
    \includegraphics[width=0.49\linewidth]{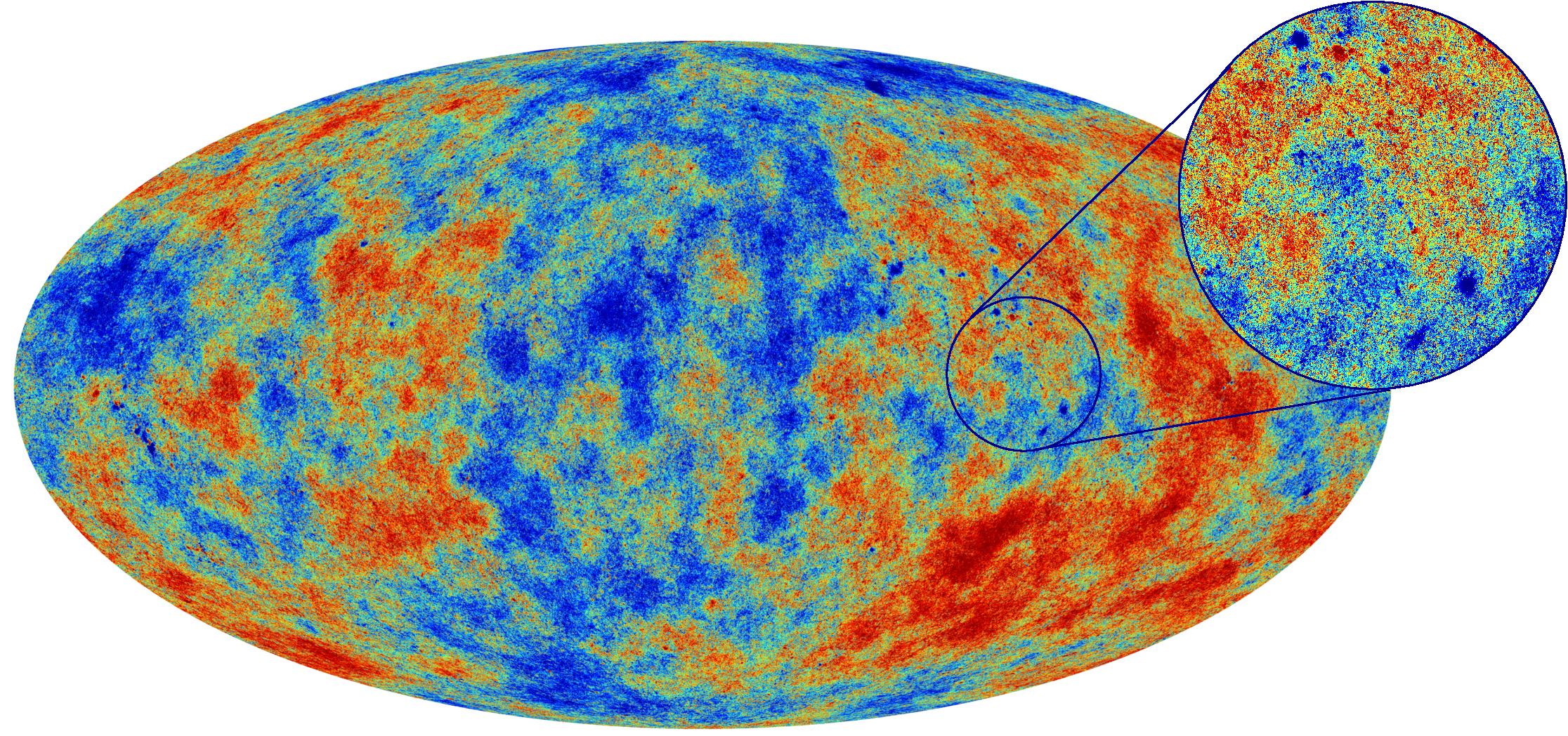}
    \includegraphics[width=0.49\linewidth]{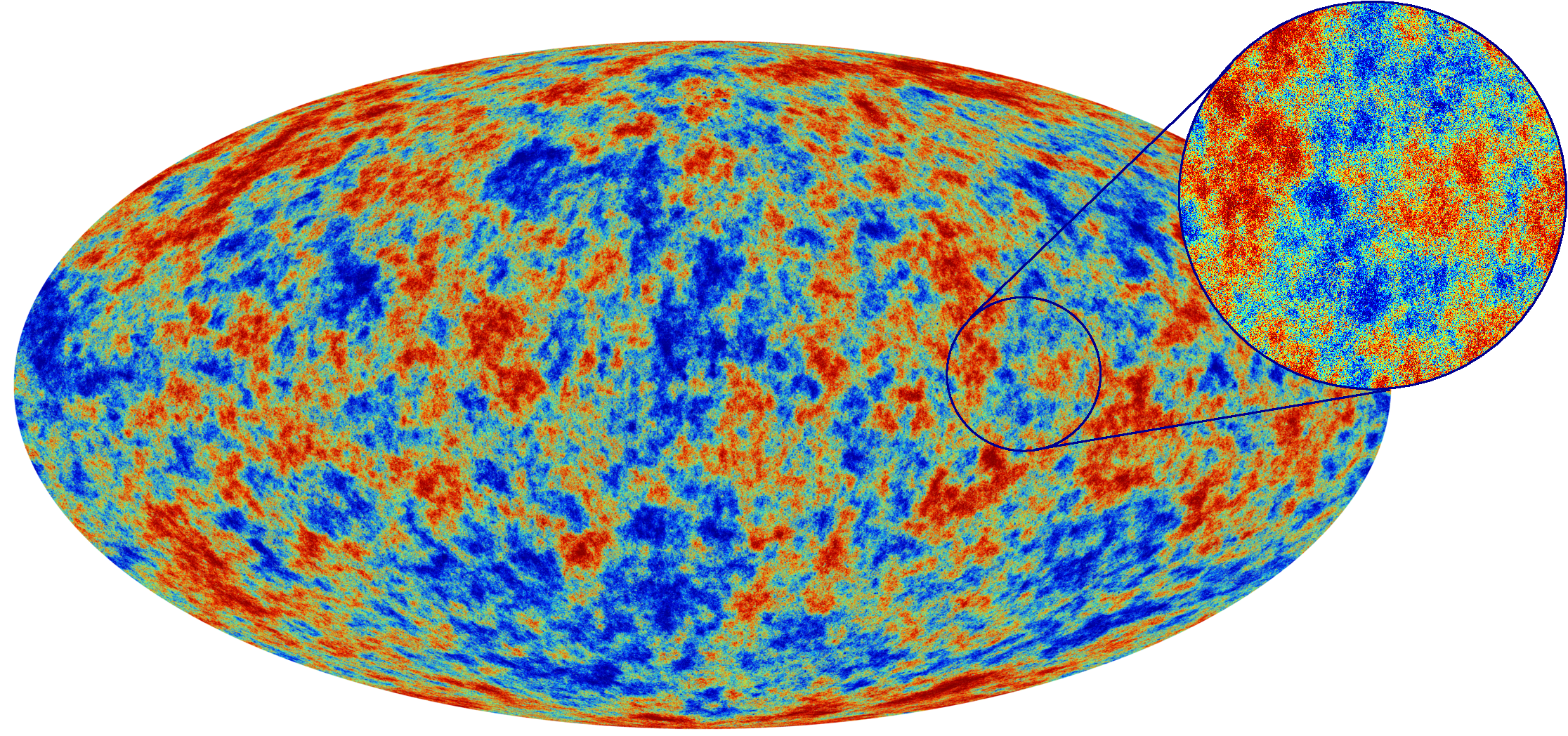}
    \caption{Full sky maps in a Mollweide projection for the Compton y parameter (see equation \ref{eq:yparam}) and the kSZ (see equation \ref{eq:kSZeffect}) are shown in the upper and lower panel, respectively. Left column is from the hydro simulation, right column is from websky. To emphasize the structures, additionally a zoom in to some region is shown and overall a color scale based on histogram equalization is used.}
    \label{fig:sim_maps}
\end{figure*}

\subsubsection{ \textit{Cosmoboost} - a harmonic-space boosting code}\textit{Cosmoboost} is a  boosting method working in harmonic space \citep{Yasini:2017jqg, Yasini_2020}.
The harmonic coefficients $a_{lm}$ of a given map 
are boosted through the computation of the boosting kernel ${}^d\mathcal{K}_{L\ell}^{m} $, as specified in \cite{Yasini:2017jqg}.
This boosting method allows to boost any type of frequency spectrum and Doppler weight, and it is therefore more general than {\it Pixell}.
It naturally incorporates both Doppler and aberration effects, and it allows maps with different Doppler weights to be boosted.
The analysis of specific intensity, partial sky maps, which are ultimately observed with future experiments, has been presented in \citep{Yasini:2017jqg} for multipoles up to $\ell = 3000$. The effects of the boosting on the temperature maps variance is addressed in \citet{Yasini_2020}.
In this work we present we updated and improved version of the code which has been developed  for a more agile handling of  small-scales. The updated code is public \footnote{For further information on \textit{Cosmoboost}, visit \url{https://github.com/maamari/Cosmoboost}}. 

\section{Simulations}\label{sec:simulations}

In order to test the effects of boosting, we need to start from unboosted simulations. 

For analyses of the CMB we use the \textit{Pixell} package to simulate full sky Gaussian CMB maps using the lensed CMB power spectrum from CAMB \citep{Lewis_1999bs}. We use the best fit \textit{Planck} cosmological parameters \citep{Planck2018VI} and generate maps in the plate carr\'ee (CAR) pixelization with a pixel resolution of 0.5 arcmin. We considering lensing and bispectrum measurements we use lensed CMB realization that are computed using \textit{Pixell} from unlensed CAMB CMB power spectra.

As for the SZ effects, we analyzed two  sets  of simulations: the WebSky simulation by \cite{Stein:2020its} and the Hydrodynamic simulation by \cite{Dolag:2015dta} - hereafter Dolag et al. These two sets of data were generated using very different methods, and extend to different redshift values, as detailed  in Sections \ref{sec:Klaus} and \ref{sec:websky}. 

\begin{figure}
    \centering
    \includegraphics[width=\linewidth]{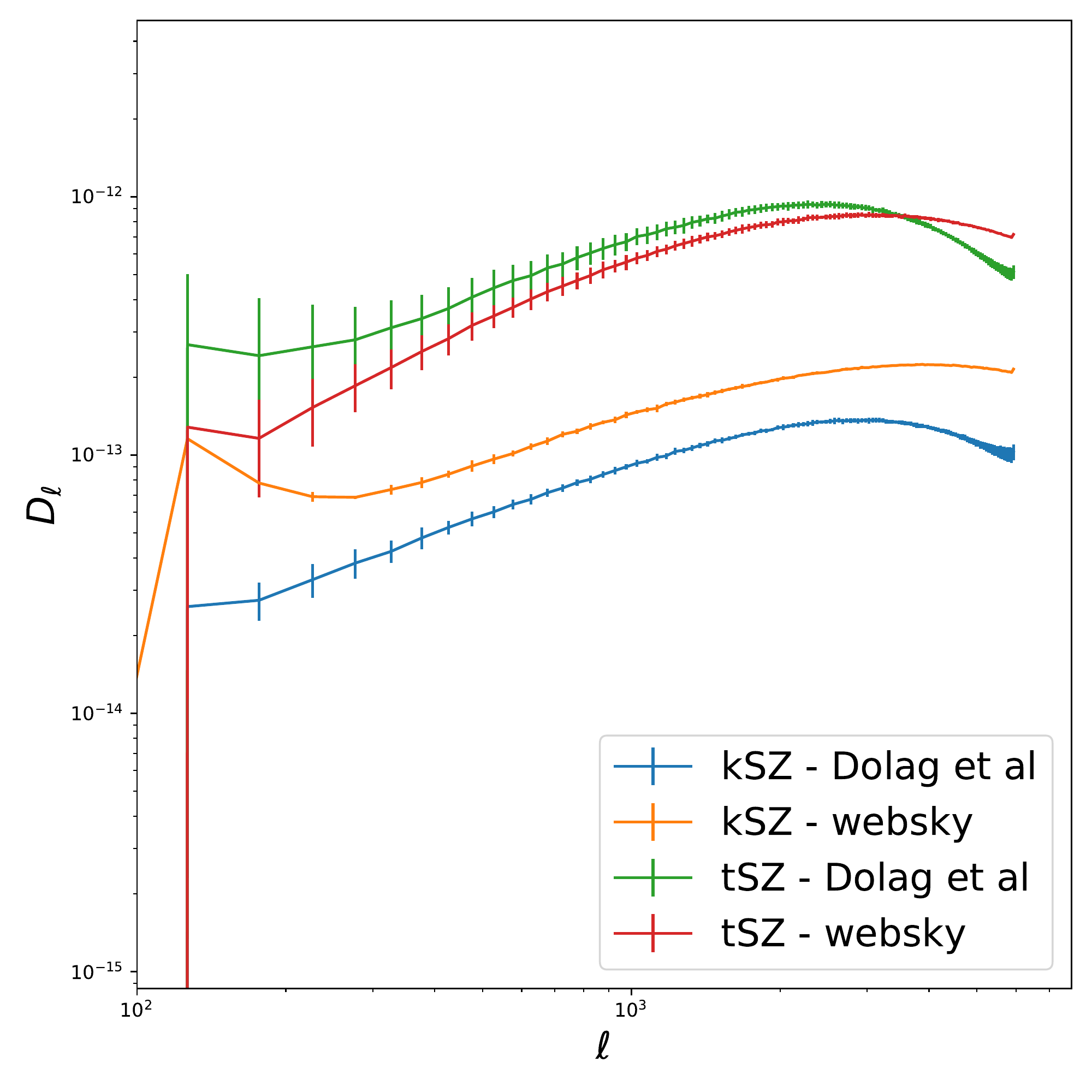}
    \caption{The pseudo-C$_\ell$ power spectrum for the  tSZ and kSZ for the Websky simulations and \citet{Dolag:2015dta} hydro-simulations. These are measured in the 12 healpix megapixels, with the error bars denoting the measured spread. The spectra are very different from each other in harmonic space; this is due to the different redshift ranges of the two simulations and is explained in further detail in Sections \ref{sec:hydroSims} and \ref{sec:webskySims}}
    \label{fig:tszksz_ws_hs}
\end{figure}

The two diverse  methods lead to significant differences in the properties of the simulations, as can be seen both in the maps (Fig.~\ref{fig:sim_maps})  and in the power spectra (Fig. ~\ref{fig:tszksz_ws_hs}). 
The Dolag et al. simulations show significantly less power in the kSZ effect at all scales. As for the tSZ, the simulations show similar overall power but different scale dependence of the power. These effects primarily arise from the different redshift ranges covered by the simulations as discussed further below.

In order to perform statistical analysis of the SZ effects, we need to analyze several independent data sets. To do this, we divide the full sky maps into 12 partial sky maps, corresponding to the 12 megapixels defined by a healpix map with nside=1 \citep{Gorski_2005}. We rotate each map (in harmonic space using the healpy Rotator.rotatealm function) such that the middle of each megapixel is in the direction of the boost, which is in this case the z hat direction, and then boost the map. We then create a mask which covers all points except the megapixel of interest, centered in the z hat direction, and add an apodization of 2 degrees (in order to negate edge effects).

\subsection{Hydrodynamical Simulation} \label{sec:hydroSims}
\label{sec:Klaus}

Cosmological, hydrodynamical simulations are starting to cover volumes 
which allow the construction of detailed full sky maps for various observables,
like the tSZ and the kSZ signal. Here, following \citet{Dolag:2015dta} we
use the so called {\it Magneticum} simulation {\it Box1/mr},
which covers a co-moving volume of almost 2.1Gpc$^3$, as well as {\it Box0/mr}, which covers a co-moving volume of 55.6Gpc$^3$.
For a detailed description of the simulations see \citet{Dolag:2015dta,2018MNRAS.478.5320S}.
In short, these simulations
cover all the important galaxy formation processes and their related feedback
processes onto the inter stellar and inter cluster medium including: star-formation
and their associated energy release through supernovae of type Ia and II, and
 the evolution of super massive black holes and their associated
AGN feedback. Also, plasma physical processes such as cooling, including
the contributions of various metal species and the presence of the CMB and
the ultraviolet (UV)/X-ray background radiation from quasars and
galaxies, as well as thermal conduction are treated in a proper way.
Thereby, the inter cluster medium in these simulations produces a tSZ signal that compares very well with the observed pressure profile in galaxy clusters
\citep{2013A&A...550A.131P,2014ApJ...794...67M,2017MNRAS.469.3069G} as well
as with the mean thermal pressure in the universe and its evolution
\cite{2021PhRvD.104h3538Y}. Despite the relatively large volume of these simulations,
creating full sky maps, as shown in Figure \ref{fig:sim_maps}, without duplication of the
simulation volume -- and thereby the structure within it -- has some limitations. Here
{\it Box1} can cover up to a redshift of $z = 0.17$ using 5 slices corresponding
to the appropriate output times, while {\it Box0} can be used to extend this up to 
a redshift of $z=0.5$ using 3 slices corresponding to the according output times.
By replicating $Box0$ twice in each spacial direction, full sky maps can be
extended to $z=1.2$ using three more slices, corresponding to the matching output
times.

In addition, the local universe contains some very prominent structures in
the form of massive galaxy clusters and super cluster regions, which
dominate the tSZ signal at small multipoles \citep[see][]{dolag2005}.
Therefore we combined the {\it Magneticum} full sky maps with the
contribution obtained from a constrained, local universe simulation
\citep[see][]{dolag2005,Dolag:2015dta} covering the redshift range of
$0 < z < 0.027$. These simulations were performed using the same galaxy
formation physics as described before for the {\it Magneticum}
simulation. Thus they recover well the observed signal of local galaxy
clusters like Coma \citep{2013A&A...554A.140P,Dolag:2015dta} or Virgo
\citep{2016A&A...596A.101P}. These local universe structures are visible in Figure \ref{fig:sim_maps} and partially explain why the Dolag et al. tSZ simulation has more power on the largest scales (small $\ell$) than the websky simulation. These features are less important for the kSZ maps.

All simulations assume a flat $\Lambda CDM$ model with slightly different
cosmological parameter. In the {\it Magneticum} simulations
$(\Omega_m, \Omega_b, \sigma_8, h, n_s) = (0.272, 0.0456, 0.809, 0.704, 0.963)$
are assumed for the cosmological parameters, whereas the {\it Local Universe}
simulation is based on $(\Omega_m, H_0, \sigma_8, h) = (0.3, 100h, 0.9, 0.7)$.
These  small differences in parameters are however not expected to impact the  conclusions  presented here.

\subsection{WebSky Simulation} \label{sec:webskySims}
\label{sec:websky}
 Instead of the computationally intensive hydrodynamic method, the WebSky realizations use the mass-Peak Patch approach \citep{Stein:2018lrh} and second order Lagrangian perturbation theory (2LPT) \citep{Bouchet:1994xp} to generate catalogs of dark matter halos and a matter field component, respectively. While these are ``approximate" methods, they have been validated extensively against more computationally expensive N-body simulations at high resolution. Using this method, the Websky cosmological realizations were able to extend to a redshift of $z = 4.6$ over the full-sky with a volume of about 600 (Gpc/h)$^{3}$ and with halos resolved down to $\sim 1 \times 10^{12} M_\odot $ \citep{Stein:2020its}. 
 
From the large scale structure realization, a map of the tSZ effect is generated by ``pasting'' spherically symmetric gas profiles onto the halos and using Eq. \ref{eq:yparam} to generate y-maps by projecting along the line-of-sight. The pressure profiles are obtained from fits to hydrodynamical simulations \citep{Battaglia_2012}.

The halo contribution to the kSZ is calculated in an analogous manner: electron density profiles are ``pasted'' onto halos and these are projected to kSZ maps using Eq. \ref{eq:kSZeffect}. The electron number density profile is also obtained from fits to hydrodynamical simulations \citep{Battaglia_2016}. This halo contribution is complemented by a contribution from the unbound field component. The electron density is assumed to be a biased tracer of the underlying dark-matter field (with $b=1$) that is then projected along the line-of-sight via Eq.\ref{eq:kSZeffect}, we refer the reader to \citet{Stein:2020its} for a more detailed description.

The WebSky simulation was generated according to a  flat $\Lambda$ CDM model with the following cosmological parameters: $(\Omega_m, \Omega_b, \sigma_8, h, \tau) = (0.31,0.049,0.81,0.965,0.68,0.055)$.

\section{Two-point statistical analysis}\label{sec:twoPoint}
Before discussing how higher-order statistics are impacted by our motion we examine how the CMB, kSZ and tSZ power spectra are impacted.
Some work has been done in this direction when considering the CMB temperature power spectrum only \citep{Challinor_2002,Jeong:2013sxy}, and we will compare with these results whenever possible to validate the \textit{Cosmoboost}  and \textit{Pixell} codes. 
Here, we aim at assessing tools to compute this effect at high $\ell$ in both temperature and polarization.
For the tSZ and kSZ effects, what we present  is both a new result and a validation of our frequency-depedent power spectrum method.

For the results of this section, we use the Namaster package \citep{Alonso_2019} to compute mask-deconvovled power spectra. When using the power spectrum boosting method (Eq. \ref{eq:Jeong} and Eq. \ref{eq:jeongExtended}) we compute the power spectrum on the full-sky and use this in our boosting formulae.

\subsection{Validation on CMB maps}

\begin{figure}
    \centering
    \includegraphics[width=\linewidth]{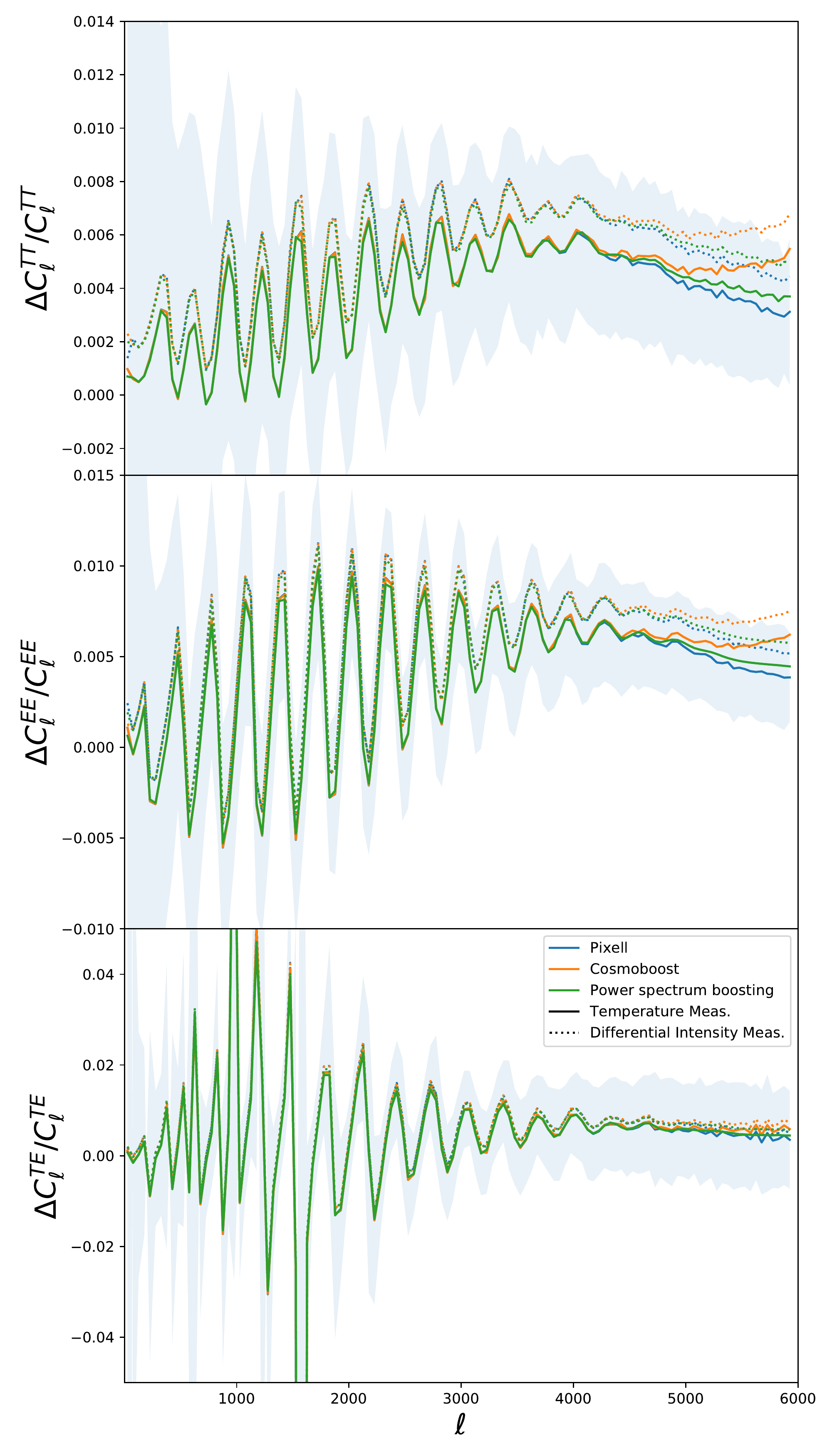}
    \caption{The fractional change between the boosted and rest-frame CMB power spectra from the half of the sky in the direction of the boost. The same 100 rest-frame maps are boosted by our three procedures as described in Section \ref{sec:boostingMethods}. The solid and dotted lines represent results for temperature and differential thermodynamic measurements at 150 GHz, respectively. The measured psuedo-Cl power spectra are binned in bins of width $\Delta \ell=50$. The shaded region is the scatter on the rest-frame power spectra from the 100 simulations. The power spectrum boosting is done via Eq. \ref{eq:Jeong} for temperature measurements and Eq. \ref{eq:jeongExtended} for differential intensity measurements. }
    \label{fig:cmb_north_ps}
\end{figure}
Here, we compare the outcomes  of the three methods discussed in Section \ref{sec:boostmet}. We also discuss the performances in terms of computational time and memory needed. We perform our analysis for temperature measurements and differential thermodynamic measurements and compare our results to the appropriate power spectrum boost formulae: either Eq.\ref{eq:Jeong} (which has been tested already up to $l \simeq 3000$) or Eq. \ref{eq:jeongExtended} (a new result of this work).

Starting with 100 Gaussian realizations of the CMB we boost each map in both real space (using \textit{Pixell} ) and harmonic  space (using \textit{Cosmoboost}).  We then mask the sky to select only the half in the direction of the boost. The mask is apodized by smoothing with a 2$^\circ$ Gaussian to minimize the impact of masking induced mode coupling \citep{Peebles_1973,Hivon2002}. To compare these methods, we calculate the binned power spectrum of both boosted maps, using Namaster with in bins of width $\Delta \ell=50$ \citep{Alonso_2019}, and determine the percent difference between the rest-frame and boosted-frame power spectrum. 
Additionally, we apply the power spectrum formula in Eq.\ref{eq:Jeong} (for $\Delta T/T_{\rm CMB}$ measurements) and  Eq. \ref{eq:jeongExtended} (for the differential thermodynamic measurements) directly to each map's rest-frame power spectrum as a further comparison.

 The results of this comparison are plotted in Figure \ref{fig:cmb_north_ps}, where we show the average spectra and the variance of the 100 realizations. We see that all methods produce highly consistent results up to $\ell_{max}=5000$ in temperature and polarization.  This is the case for both $\Delta T/T_{\rm{CMB}}$ measurements (compared with Eq.\ref{eq:Jeong})  and for differential thermodynamic measurements at 150 GHz (compared with the newly derived formula \ref{eq:jeongExtended}). Note the spikes in the TE power spectra arise as the denominator crosses zero.  The impact of the boost on differential thermodynamic measurements is $\sim 30\%$ larger than for temperature measurements, as the modulation term is increased. These result provide the first validation of our frequency dependent power spectrum boosting formula,  Eq. \ref{eq:jeongExtended}: the good agreement between this method and the two other methods provides validation for the accuracy of this method. The difference seen between the $\Delta T/T_{\rm{CMB}}$  measurements and the differential thermodynamic measurements highlights the need for this formula; current CMB observation measure the later quantity while theory codes compute the former and our formula provides a fast and accurate method to map between the two.   
  
 At smaller scales we see  small discrepancies. 
 These may be linked to numerical effects in one or the other methods. In principle, the \textit{Cosmoboost} formula is exact and we have checked that the range extent of the kernel is not impacting the result.
As described in Section \ref{sec:boostmet} the \textit{Pixell} method uses an interpolation step to evaluate the aberration. This interpolation step introduces a transfer function, akin to the pixel window function, that damps the small scale power (see \citet{Yoho:2012am} for further discussion). This could partially explain the different behaviour seen at high $\ell$, however it is unlikely that this completely explains the difference as the estimated transfer function is smaller then the observed difference. The power spectrum boosting formula is only correct to first order in $\beta$ which could lead to an inaccurate prediction on the smallest scales.
Thus we do not have a conclusive interpretation for these small discrepancies.

\begin{table*}
\begin{tabular}{l| cccc|cc|} 
 $\ell_{\mathrm{max}}$ & \multicolumn{4}{c}{CPU  usage (core-mins)}  &  \multicolumn{2}{c}{Memory Usage (GB)} 
\\ \hline
& \textit{Cosmoboost} (kernel) & \textit{Cosmoboost} (boost) & \textit{Pixell} (precomputation) & \textit{Pixell} (boost)  & \textit{Cosmoboost} & \textit{Pixell}  \\
                           \hline\hline 
1000 & 0.2 & 0.01 & 1 & 1.6& 0.03 & 2.7 \\
2000 & 0.8 & 0.02 & 2.5 & 0.1 & 6.8 & 15   \\
4000 & 8.9 & 0.1 & 10 & 0.5 & 39 & 65 \\
6000 & 83 & 0.5 & 24 & 2.5 & 120 & 160  \\
\end{tabular}
\caption{ Resource consumption for the \textit{Cosmoboost} and \textit{Pixell} codes to boost a single temperature map as a function of $\ell_{max}$. Both codes can run in two stages: a one-off computation, for \textit{Cosmoboost} this is the computation a kernel and for \textit{Pixell} the evaluation of the interpolation locations, and then the use of these components to a boost specific realization. As can be seen for boosting a single map the \textit{Pixell} code is faster, at the cost of a larger memory footprint. However \textit{Cosmoboost} is significantly faster when boosting multiple maps. }
\label{tab:codeResources}
\end{table*}
We then assessed the computational resources needed to boost the map.
Table~\ref{tab:codeResources} shows the resources needed to create boosted maps with \textit{Cosmoboost} and  \textit{Pixell}. Both codes have the option to reuse parts of the computation. This means boosting of multiple maps can be significantly faster than a single map: in \textit{Cosmoboost} the kernels can be resued and in \textit{Pixell} the interpolation locations can be resused.
When using \textit{Cosmoboost} to boost the spherical harmonic coefficients of a temperature map up to $\ell = 6000$, roughly $120$ gigabytes of memory and $82$ minutes of runtime are required to generate a kernel of width $\Delta \ell = 20$, which is adequate for boosting temperature and polarization spectra up to $\ell=6000$, and the half a minute of runtime  is required to boost a map with these kernels. For further information on $\Delta \ell$ selection, please visit the \textit{Cosmoboost} repository. For \textit{Pixell} the  precomputation resources are $24$ core minutes and $160$ GB of memory and the boosting time is about 2.5 minutes for a temperature map. For repeated boosting the resource requirement is just the boosting operation, which is significantly faster for both codes, with \textit{Cosmoboost} being several times faster.   
It is possible, though out of the scopes of this work,  to optimize the efficiency of \textit{Cosmoboost}  by altering the dimensions of the kernel and adapting it to the $\ell$ value under examination. The boosting runtime for temperature and polarization maps are approximately triple the temperature alone requirements.

\begin{figure}
    \centering
    \includegraphics[width=\linewidth]{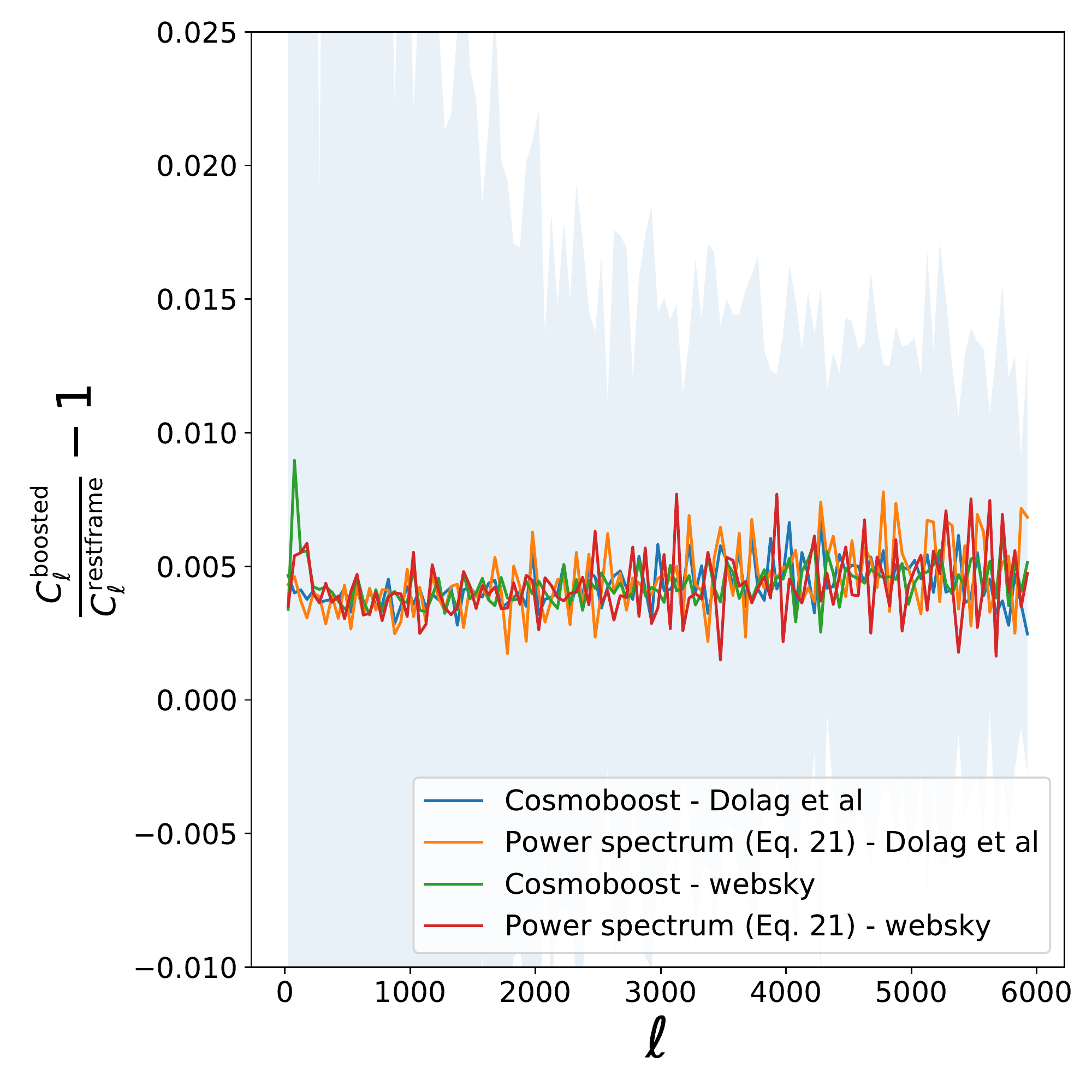}
    \caption{The ratio of the boosted and rest-frame kSZ power spectra for the Websky and Dolag et simulations for differential thermodynamic measurements at 143 GHz. The results are the average for boosts on the 12 healpix megapixel patches, as described in the text, and the shaded region is the sample variance from the 12 rest-frame patches of the Websky simulations.} \label{fig:kSZ_healpix_ps}
\end{figure}

\begin{figure}
    \centering
    \includegraphics[width=\linewidth]{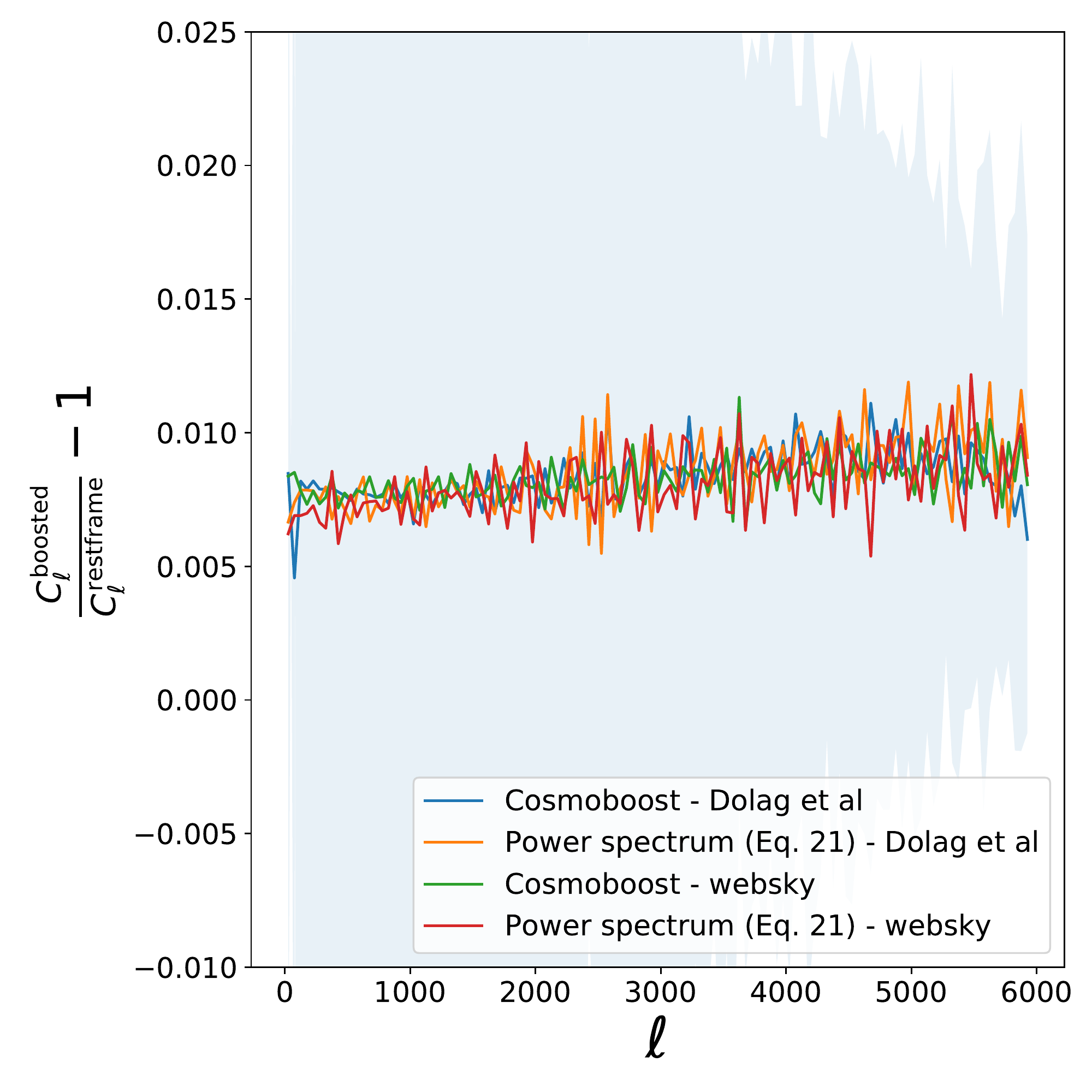}
    \caption{The ratio of the boosted and rest-frame tSZ power spectra for the Websky and Dolag et simulations  for differential thermodynamic measurements at 143 GHz. The setup is otherwise identical to that used for the kSZ effect shown in Fig \ref{fig:kSZ_healpix_ps}. The blue region denotes the sample variance of the rest-frame power spectrum across the 12 patches.} \label{fig:tSZ_healpix_ps}
\end{figure}

\begin{figure}
    \centering
    \includegraphics[width=\linewidth]{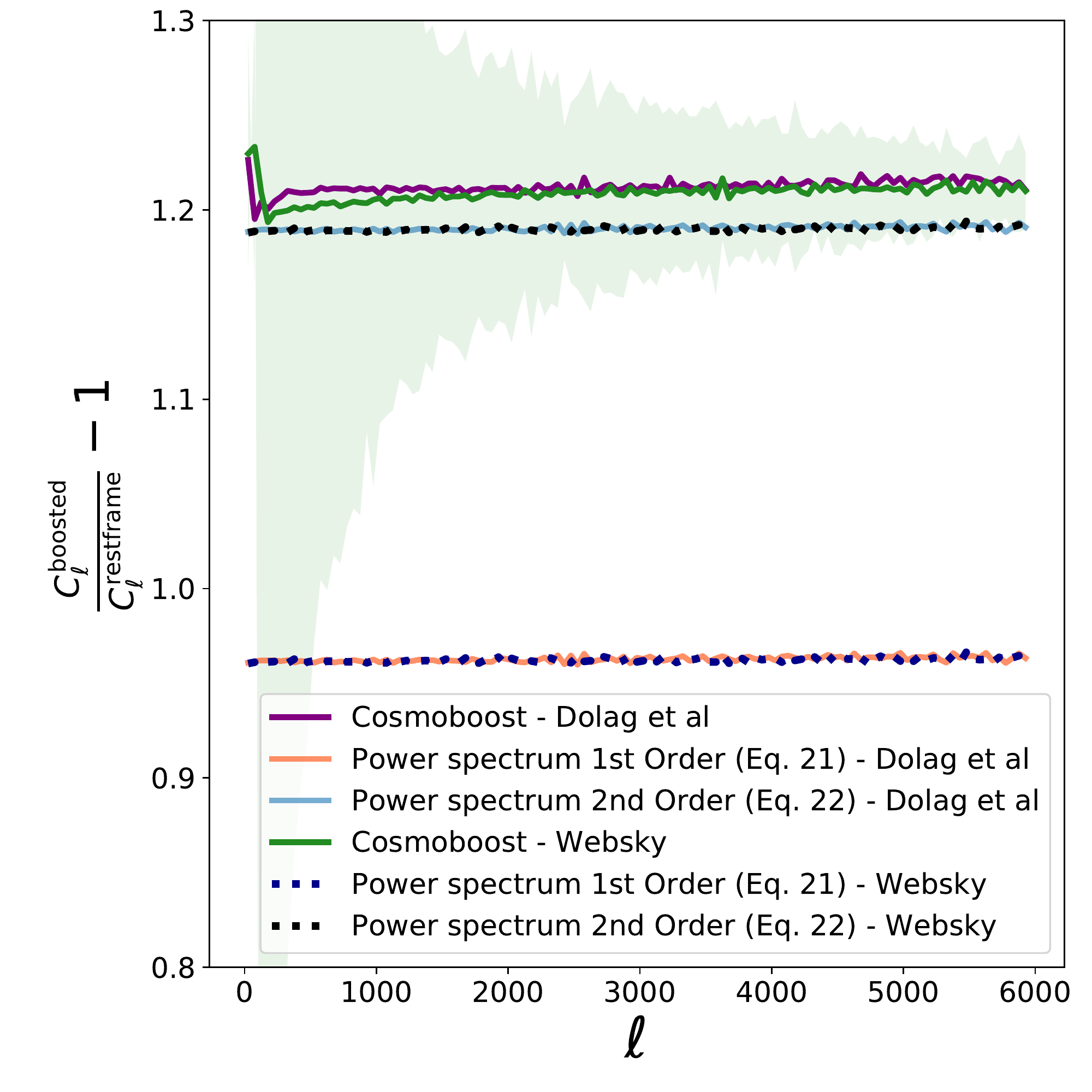}
    \caption{The ratio of the boosted-frame and rest-frame tSZ power spectra at 217GHz using the Dolag et al and Websky simulations. The experimental setup is otherwise identical to Fig. \ref{fig:tSZ_healpix_ps}. At this frequency we find that higher order Doppler terms are necessary to accurately describe the boosting effects. We provide an analytic formula for these terms in Eq. \ref{eq:jeongExtendedWith2ndFreq}. Note that whilst the fractional effect is large, at 217 GHz the tSZ signal is close to its null and so the absolute correction is still small.}
    \label{fig:tSZ_217GHz}
\end{figure}

\subsection{Analysis of kinetic Sunyaev Zel'dovich maps} \label{sec:powtsz}
The kSZ effect has the same frequency response as the primary CMB signal, however it has a significantly different $\ell$ dependence. This means that the expected impact of boosting effects will be different from the effects on the primary CMB. In Figure \ref{fig:kSZ_healpix_ps} we show the results of boosting the kSZ maps using the \textit{Cosmoboost} on the 12 megapixel patches for both simulations (given the good agreement between \textit{Cosmoboost} and \textit{Pixell} for primary CMB anisotropies we focus on just \textit{Cosmoboost} here for clarity).  We see that the effect of boosting is primarily at 0.5\% increase in the kSZ power on all scales and the impact of the boosting effects is similar in both the websky and Dolag et al simulations. Additionally, the blue band shows the sample variance computed with the rest-frame patches; this significantly exceeds the size of the Doppler effects on almost all scales, and can reach a few percent. Note that the sample variance computed in the boosted frame and the rest frame are statistically consistent and that the scatter between the rest-frame and boosted-frame measurements is smaller than the sample variance by more than a factor of five on all scales. These results suggest that boosting has an insignificant effect on kSZ power spectrum measurements.

We also show the results from the power spectrum boosting formula. It shows very good agreement with \textit{Cosmoboost} providing further validation of our formula (Eq.\ref{eq:jeongExtended}). Note that we observe significant scatter between the \textit{Cosmoboost} and power spectrum boosting results. This arises as the \textit{Cosmoboost} results are computed on the 12 megapixel patches whereas the power spectrum formula is evaluated using the full-sky power spectrum. We cannot use the power spectrum measured in the patch as this will be convolved with the mask \citep{Hivon2002}, which does not commute with the boosting operation.

\subsection{Analysis of thermal Sunyaev Zel'dovich maps} 
We then perform a similar analysis of the tSZ maps. Note that currently the \textit{Pixell} code can only boost maps of signals with CMB like frequency dependence, so all maps were boosted using the \textit{Cosmoboost} code.

In Figure \ref{fig:tSZ_healpix_ps} we plot the fractional change on the tSZ power spectrum using the 12 healpix megapixel patches for the two types of simulations at 143 GHz. Firstly we see that our power spectrum formula shows very good agreement with \textit{Cosmoboost}. Second we see that the size of the boosted correction is very similar for the two simulations.
In both cases, the average of the effect is at most 1\% on all scales. As in the case of the kSZ power spectra, the sample variance exceeds the size of the boosting effect on all scales and the scatter between the rest-frame and boosted-frame power spectra is significantly smaller than the sample variance.

The tSZ spectral response, relative the the CMB anisotropies, has a null near 217GHz (note the singal is not exactly null at 217 GHz). Around these frequencies higher order Doppler terms become important. This is seen in Fig. \ref{fig:tSZ_217GHz} where the first order power spectrum boosting formula, Eq. \ref{eq:jeongExtended},   is still highly inaccurate, despite providing an O(1) correction to the signal. We find that second order derivatives of the frequency response, captured by our second order formula, Eq. \ref{eq:jeongExtendedWith2ndFreq}, are very significant and provide a much more accurate description of the modulation effect. We note that these large relative corrections are primarily due to the fact that the tSZ signal is small near the null, rather than these corrections being large in absolute terms - at 217 GHz the tSZ power spectrum is suppressed by a factor of $5\times 10^{-5}$ compared to the tSZ power spectrum at 143 GHz. The importance of this effect for experimental measurements depends on the details of the instrument bandpass.  For example when integrated against the ACT bandpass \citep{Marsden_2014}, the boosting induced correction to the tSZ power spectrum in the 217 GHz is reduced to $\sim 4\%$. Measurements at 217GHz are useful for isolating or removing the tSZ signal from CMB maps as the signal is close to null. To ensure there is no bias from this effect it may be important to account for motion induced shifts.

\section{One-point Statistical Analysis } \label{sec:onePointAn}

\begin{figure}
    \centering
    \includegraphics[width=\linewidth]{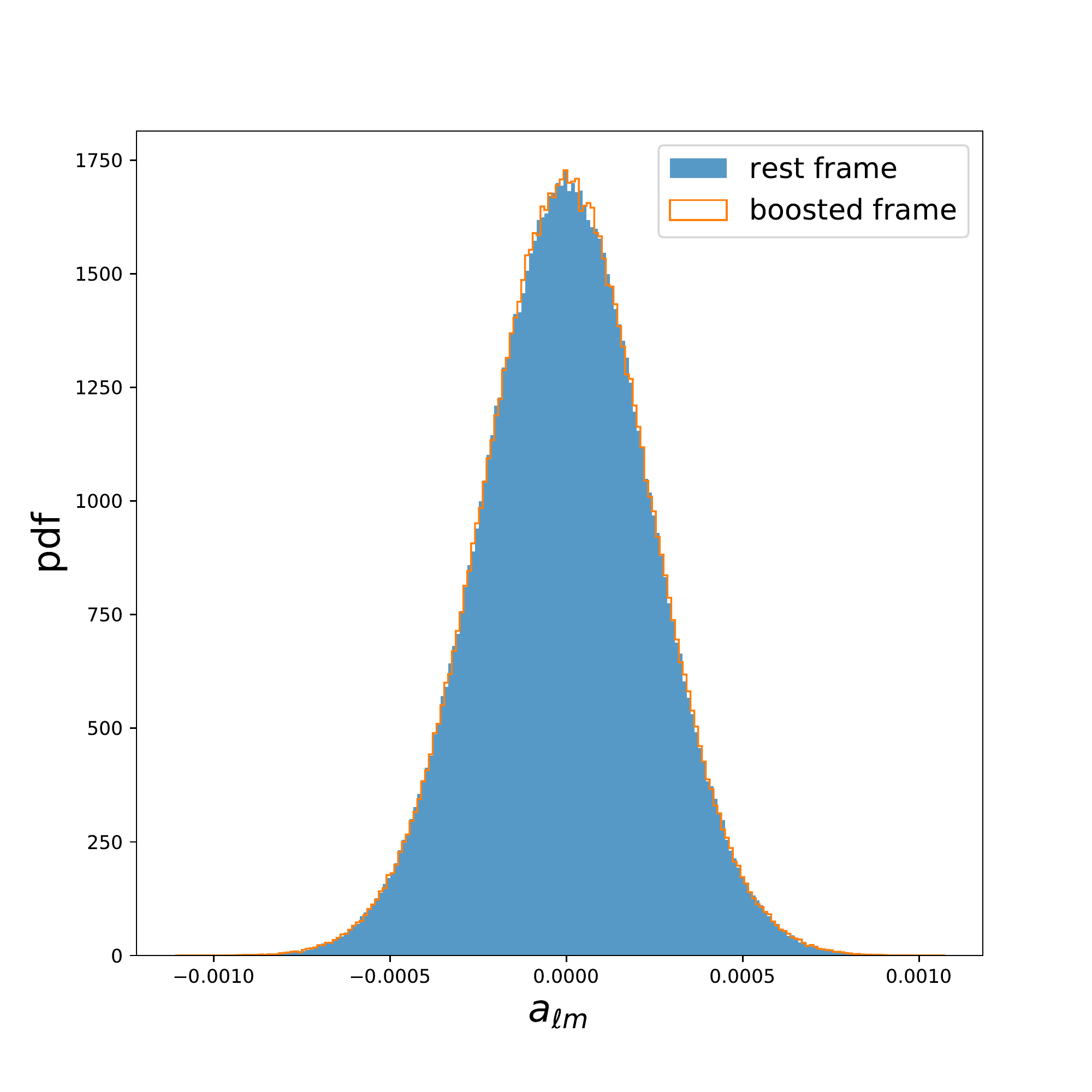}
    \caption{The one-point pdf of rest-frame and boosted-frame CMB $a_{\ell m}$ coefficients at $\ell=5000$ for 100 simulations. We measure the $a_{\ell m}$ coefficients after applying a mask that retains the $50\%$ of the sky in the direction of the boost. We plot the distribution of both the real and imaginary parts as well as all the m modes together.}
    \label{fig:onePdf}
\end{figure}

The most complete one-point statistic is the one-point probability density function (pdf), which is our case is the distribution of each of the $a_{\ell m}$ coefficients. We examined these distributions for all the simulations and an example of this is shown in Figure \ref{fig:onePdf}. 
Instead of summarizing
the information in the plethora of pdfs we instead focus on a set of compressed one point statistics: the third and fourth standardized moments. The third moment, also known as the skewness - $\hat{\mu}_3$, is defined as
\begin{align}
    \hat{\mu}_3 = E\left[\left(\frac{(X-\mu)}{\sigma}\right)^3 \right],
\end{align}
where $\mu$ and $\sigma$ are the distribution's mean and standard deviation. The skewness is zero for a perfectly Gaussian distribution. The fourth standardized moment, the kurtosis - $\hat{\mu}_4$,is similarly defined as
\begin{align}
    \hat{\mu}_4 = E\left[\left(\frac{(X-\mu)}{\sigma}\right)^4 \right].
\end{align}
The kurtosis is equal to three for a perfectly Gaussian distribution, but from here on kurtosis will refer to Pearson's kurtosis, for which three has been subtracted (so a kurtosis of zero corresponds to a perfectly Gaussian distribution). To quantify the non-Gaussianity of an $a_{lm}$ distribution, we calculate the deviation of skewness and kurtosis from zero, in both the rest frame and the boosted frame, at each $\ell$ value. 

\subsection{CMB}
\begin{figure}
    \centering
    \includegraphics[width=\linewidth]{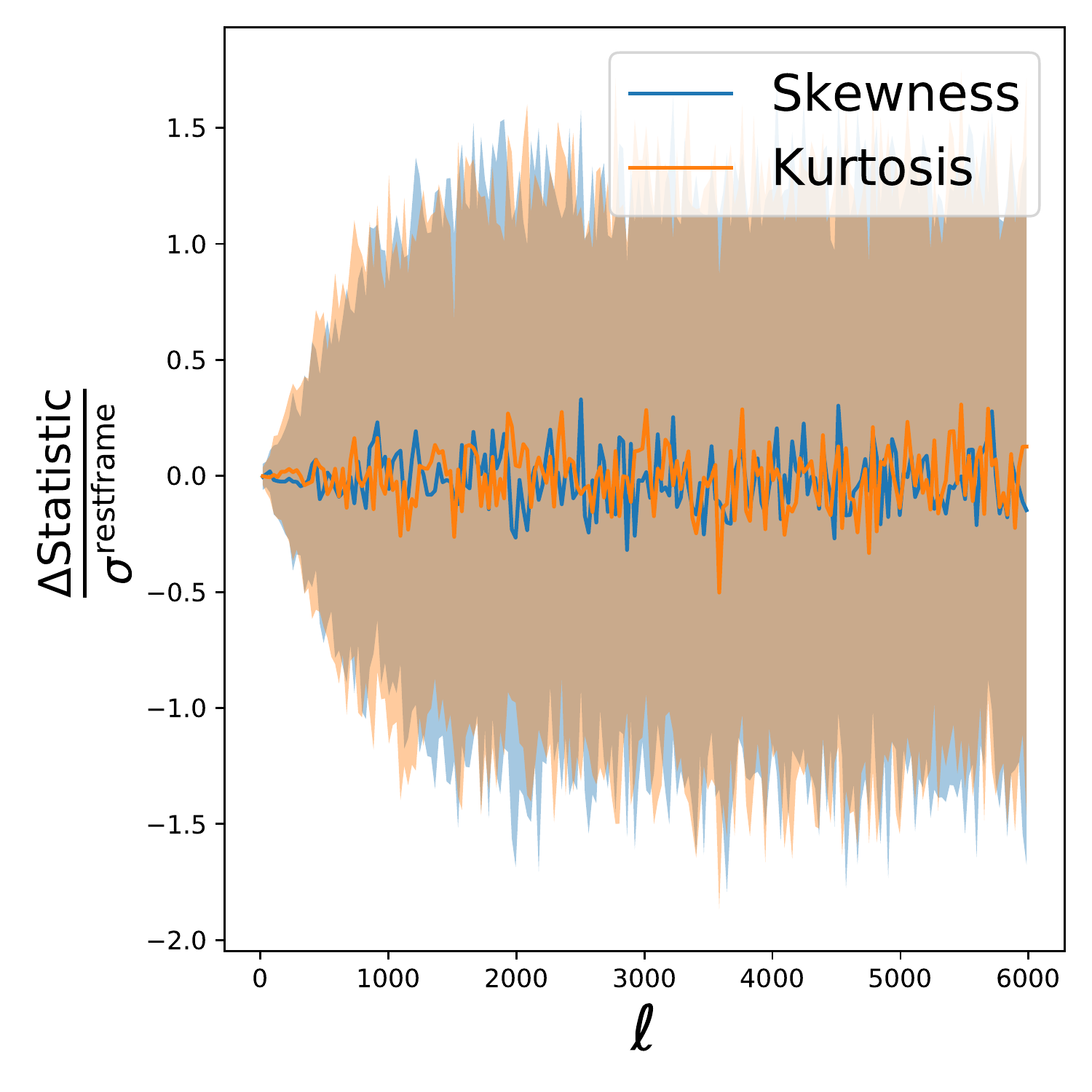}
    \caption{The change in the CMB a$_{\ell m}$ skewness and kurtosis, as a fraction of the rest-frame statistic standard deviation, induced by our motion for measurements of $\Delta T/T_{\rm{CMB}}$. We compute this using 100 CMB simulations boosted with the \textit{Cosmoboost} code. We apply a mask that retains the 50\% of the sky in the direction of the boost. The shaded regions are the measured spread of the difference. Note that the variance of the boosted-frame and rest-frame maps are statistically the same and thus the large spread of the difference indicates decorrelation and not a source of extra noise.  }
    \label{fig:cmb_hist}
\end{figure}

We boost 100 CMB realisations using \textit{Cosmoboost} and consider temperature measurements, i.e. frequency independent measurements of $\Delta T/T_{\rm{CMB}}$. In Fig. \ref{fig:cmb_hist} we show the change in the skewness and the kurtosis induced by the boost, as measured on the 50\% of the sky in the direction of our motion, compared with the standard deviation of the statistic as measured in the rest frame. First we see that, on average, the Doppler boost does not introduce any statistically detectable skewness or kurtosis. Second we see that the variance of the difference between the boosted and rest-frame measurements is comparable to the sample variance of the statistic. However the sample variance of the boosted statistic is still the same as the sample variance in the rest frame - there is no additional variance from the boosting. The scatter in the difference is thus not a sign of increased statistic variance but rather indicates an effective decorrelation between the maps. As an analogy consider a simpler case: compare the statistics of a simulated CMB map with the same map after performing a rotation. Under a rotation the $a_{\ell m}$ coefficients mix together and thus one would observe a large scatter in skewness and kurtosis measurements between the rotated and unrotated maps however the ensemble averages of the two statistics would be the same.  (This is not true for rotationally invariant statistics such as the power spectrum or bispectrum). A similar effect occurs here, the aberration mixes the $a_{\ell m}$ coefficients resulting in a scatter between the rest and boost frame measurements without producing significant changes in the ensemble average statistics.

\subsection {Sunyaev--Zeldovich effects} \label{sec:onePointSZ}

\begin{figure}
    \centering
    \includegraphics[width=\linewidth]{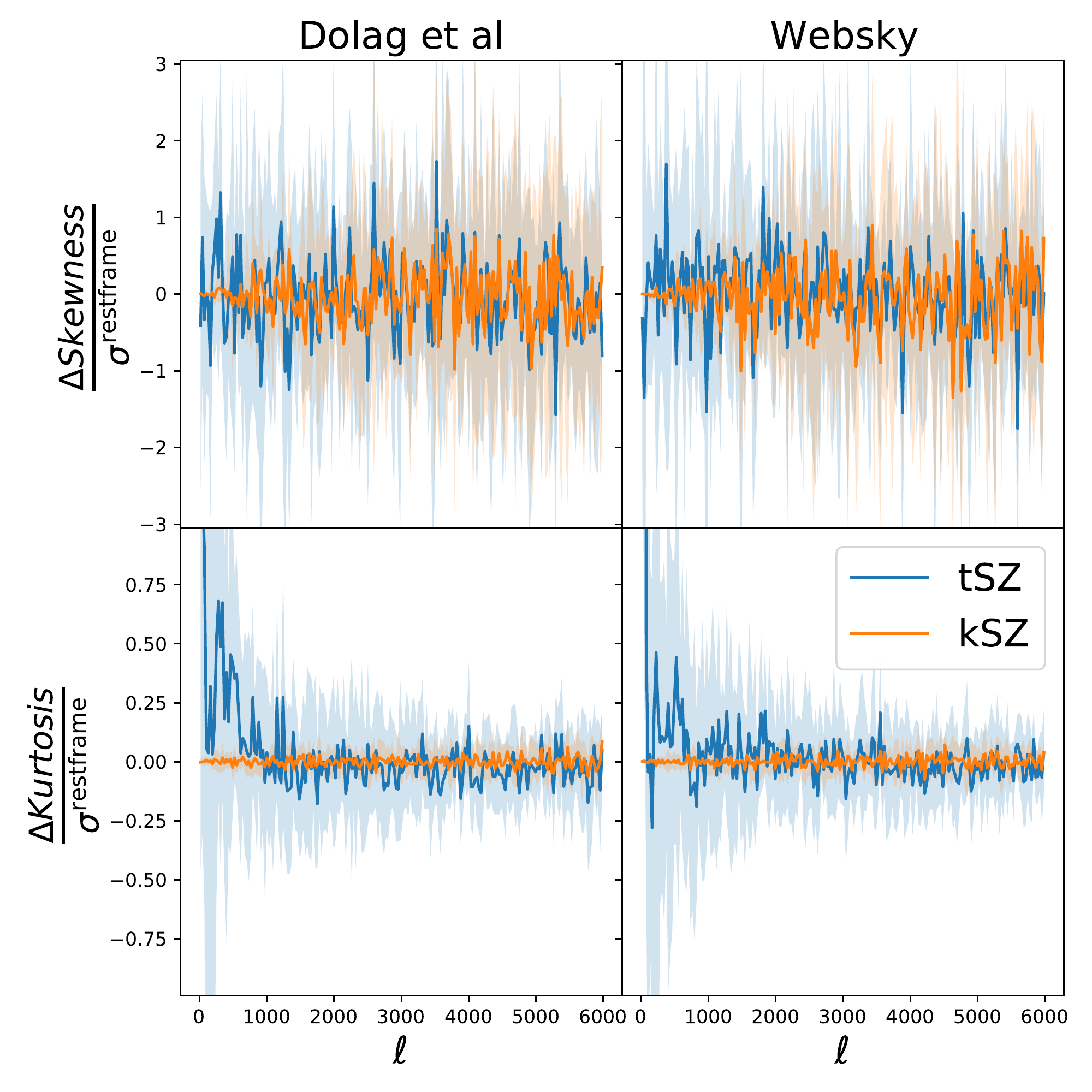}
    \caption{
    The boosting induced change in the skewness and kurtosis of the a$_{\ell m}$ coefficients from the tSZ and kSZ maps. The statistics are computed by using twelve megapixel regions, each boosted with the boost direction aligned with the patch center. The measurements are in differential thermodynamic units at 143 GHz. The shaded regions are the measured standard deviation in the difference.}
    \label{fig:SZ_hist}
\end{figure}

To investigate the impact of Doppler boosting on the Sunyaev--Zeldovich effects we make use of the 12 healpix patches of the \citet{Dolag:2015dta} and websky simulations. For both the tSZ and kSZ effects we boost the maps with \textit{Cosmoboost} and consider linearized differential thermodynamic measurements at 143 GHz. Each patch is boosted with a boost direction aligned with the center of the patch. In Fig. \ref{fig:SZ_hist} we see that the Doppler effects do not induce significant skewness or kurtosis for the kSZ effect. However we do see a small, and statistically not detectable, change in the tSZ kurtosis. We see broadly consistent results between the  Dolag et al and websky simulations. As in the case of the CMB we see significant scatter between the rest-frame and boosted-frame measurements, though as we only have twelve patches of the same sky the errors are rough estimates.

\section{3-POINT STATISTICS}\label{sec:threePointAn}
Three-point statistics probe the correlations between three spatial points or three different harmonic modes. The bispectrum, the harmonic space equivalent of the three point function, has been extensively used in cosmology to probe deviations from Gaussianity and to extract information beyond the power spectrum \citep{planck2018_IX,Smith_2009,Gil_Mar_n_2015}. In this section we consider how the Doppler and aberration effects impact bispectrum measurements. This work builds on the results of \citet{Catena_2013b}, who explored the impact of these effects on constraints on primordial non-Gaussianity using the bispectrum. \citet{Catena_2013b} find two results: firstly Doppler and aberration effects do not introduce a primordial bispectrum signal in Gaussian sky maps. Second, they observe a scatter, at the level of $\sim 30\%$ of the primordial bispectrum amplitude, between the rest-frame and boosted-frame measurements. Their results raise two questions: is the scatter between the unboosted and boosted bispectrum measurements an additional source of noise that needs to be accounted for in future studies, and what is the effect of boosting on non-zero bispectra? Our work answers these two questions. First we reexamine how primordial bispectrum estimators are affected by the Doppler and aberration terms. We include polarization data and smaller scales to investigate the scatter seen in \citet{Catena_2013b}. We also perform our analysis for differential thermodynamic measurements (whereas  \citet{Catena_2013b} considered temperature measurements), finding that this distinction leads to a systematic, but unobservable, difference. Second we consider how non-zero bispectra, specifically the ISW lensing bispectrum and the tSZ and kSZ bispectra, are impacted.

We begin this section with a review of bispectrum and primordial non-Gaussianity estimators. Then we consider how boosting effects primordial non-Gaussianity estimators applied to CMB simulations. Finally we study how the bispectrum of the  ISW-lensing, tSZ and kSZ effects are impacted by the Doppler and aberration effects.
\subsection{Overview of bispectrum and primordial non-Gaussianity estimators}
The bispectrum is defined as \citep{Spergel_1999}
\begin{align}
    B^{X_1,X_2,X_3}_{\ell_1,\ell_2,\ell_3,m_1,m_2,m_3} = \langle a^{X_1}_{\ell_1,m_1} a^{X_2}_{\ell_2,m_2} a^{X_3}_{\ell_3,m_3} \rangle,
\end{align}
where $X_i$ denotes the type of field (e.g. temperature, polarization etc). Estimation of the full-bispectrum is computational prohibitive and, for the majority of cosmological fields, any individual configuration of the bispectrum will be noise dominated. To overcome these issues a series of methods have been developed \citep{Komatsu_2005,Bucher_2010,Fergusson_2012}. These approaches utilize two techniques: first if we assume that the fields under study are parity even and are generated by homogeneous and isotropic processes the the bispectrum can be written as
\begin{align}
    \langle a^{X_1}_{\ell_1,m_1} a^{X_2}_{\ell_2,m_2} a^{X_3}_{\ell_3,m_3} \rangle =  \mathcal{G}^{m_1 m_2 m_3}_{\ell_1 \ell_2 \ell_3} b^{X_1 X_2 X_3}_{\ell_1 \ell_2 \ell_3},
\end{align}
where $ \mathcal{G}^{m_1 m_2 m_3}_{\ell_1 \ell_2 \ell_3}$ is the Gaunt integral and is a geometric factor and $b^{X_1 X_2 X_3}_{\ell_1 \ell_2 \ell_3}$ is the reduced bispectrum and contains all the physical information. Second we note that the following form of the Gaunt integral allows for the geometrical factor to be enforced by applying spherical harmonic transforms
\begin{align}
   \mathcal{G}^{m_1 m_2 m_3}_{\ell_1 \ell_2 \ell_3}   = \int \mathrm{d}^2\Omega Y_{\ell_1m_1}(\mathbf{n})Y_{\ell_,m_2}(\mathbf{n})Y_{\ell_3m_3}(\mathbf{n}).
\end{align}
In this work we use a binned bispectrum estimator as developed in \citet{Bucher_2010,bucher_2016}. The binned estimator provides estimates of the reduced bispectrum averaged over a range of $\ell$ and can be efficiently computed as
\begin{align} \label{eq:binnedEstimator}
&    \hat{b}^{X_1X_2X_3}_{i,j,k}  = \sum\limits_{\mathrm{All}\, \ell m} \frac{1}{N_{ijk}}  \int \mathrm{d}^2\Omega W^i_{\ell_1} a_{\ell_1,m_1 }W^j_{\ell_2} a_{\ell_2,m_2} W^k_{\ell_3} a_{\ell_3,m_3} \nonumber \\ & - \langle W^i_{\ell_1} a_{\ell_1,m_1 }W^j_{\ell_2} a_{\ell_2,m_2}  \rangle W^k_{\ell_3} a_{\ell_3,m_3} + \text{two permutations},
\end{align} where $N_{ijk}$ is the number of configurations summed over and $W^{i}_{\ell}$ defines the boundaries of the bins. Our implementation is described in more detail in \citet{Coulton_2019}. By averaging over several nearby configurations the binned estimator increases the signal to noise and reduces the computational complexity of the estimator. \citet{Babich_2005} and \citet{Creminelli2006} showed that bispectrum estimators require the terms in the second line of Eq. \ref{eq:binnedEstimator} in order to be optimal, unbiased estimators -  i.e. estimators that saturate the Cramer-Rao bound. These terms are known as the linear terms and we perform an analysis both with and without including them.

 If the structure of the non-Gaussianity is known a-priori, small deviations from Gaussianity are most effectively probed with template based estimators \citep{Komatsu_2005,Fergusson_2012}. We fit templates to the binned bispectrum measurements and report constraints on the amplitudes of these templates, known as $f_{\mathrm{NL}}$. Specifically the amplitudes are computed as
\begin{align}
\hat{f}_{\mathrm{NL}} = \frac{1}{6 N} \sum \limits_{i,j,k} b^{\mathrm{theory},X}_{ijk} {\Sigma^{XY}_{ijk}}^{-1} \hat{b}^{Y}_{ijk},
\end{align}
where $b^{\mathrm{theory},X}_{ijk}$ is the binned theoretical bispectrum, $\Sigma^{XY}_{ijk}$ is the binned bispectrum covariance where $X,Y$ represents the bispectrum configuration of fields, and N is a normalisation constant such that the estimator has unit response to the theoretical bispectrum. We use estimates of $f_{\mathrm{NL}}$ parameters when discussing primordial non-Gaussianity and lensing-ISW bispectra as experimental bounds have shown that these types of non-Gaussianity are small  \citep{Lewis_2011,planck2018_IX,Smith_2009,Bennett_2013}.

\subsection{Primordial bispectra}

\begin{figure}
    \centering
    \includegraphics[width=\linewidth]{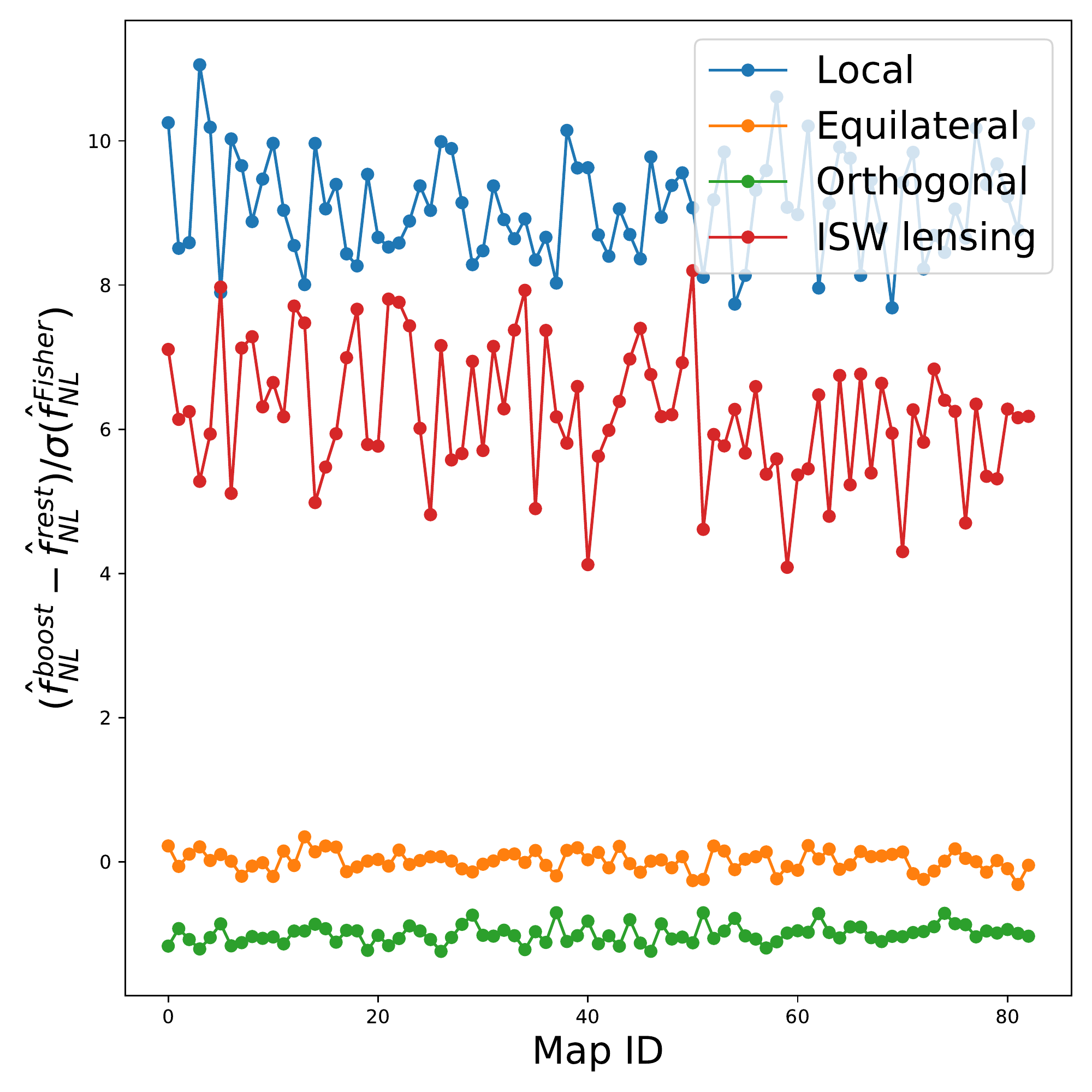}
    \caption{The different between $f_{NL}$ estimates obtained from a set of boosted and rest-frame, noiseless CMB maps. We apply a mask that retains the region of sky within $60^\circ$ of the boost direction. We do not include the linear term in our $f_{NL}$ estimator.  We normalise the estimate by the estimator Fisher variance. We see a systematic difference between the boosted and rest-frame  $f_{NL}$  values for shapes with a strong squeezed component.  However as these estimates exclude the linear term the estimator variance is significantly larger than the Fisher variance and these biases are unobservable.  \label{fig:fnl_noLiinear}}
\end{figure}

\begin{figure}
    \centering
    \includegraphics[width=\linewidth]{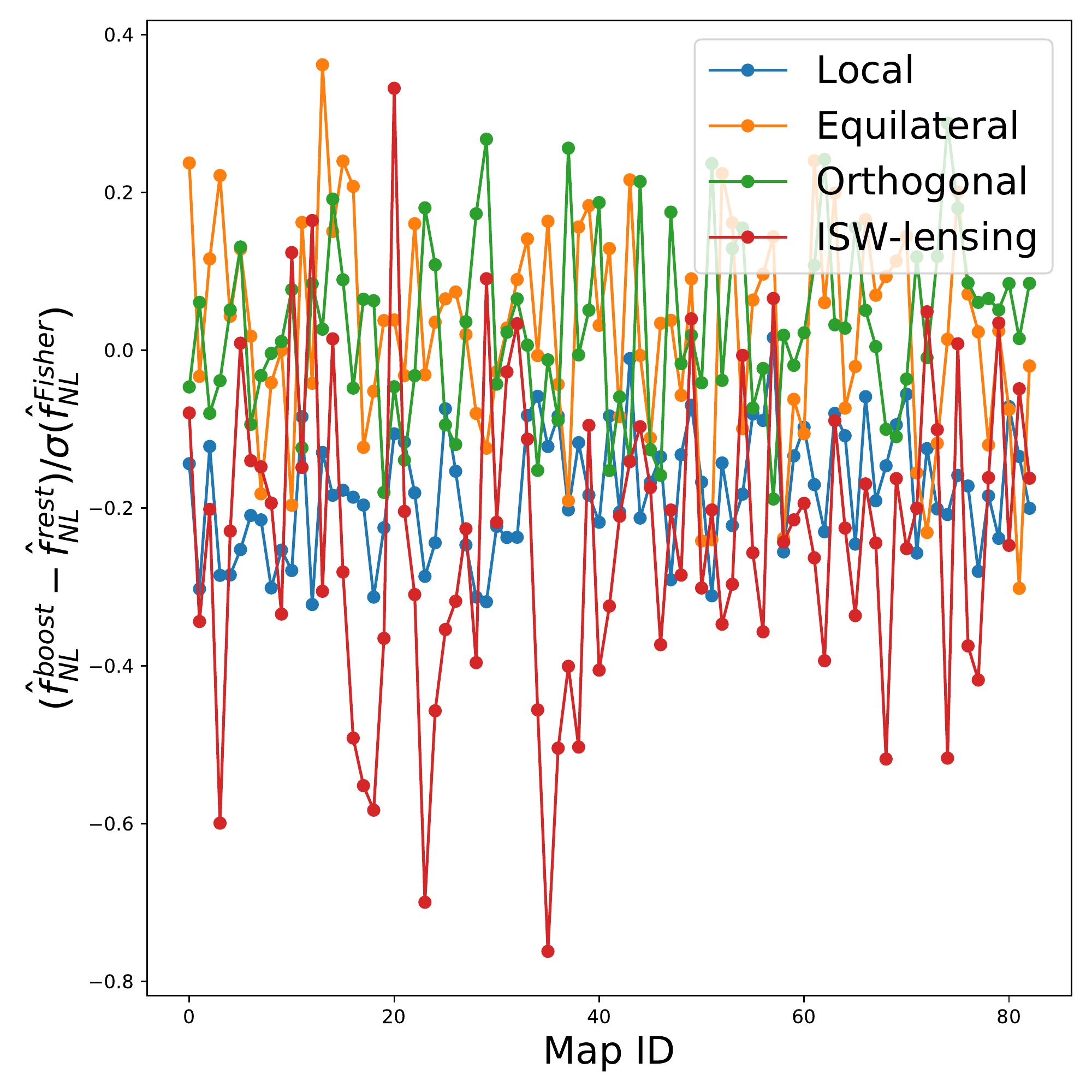}
    \caption{The same as Figure \ref{fig:fnl_noLiinear} with the inclusion of the linear term in our $f_{NL}$ estimator.  We normalise the estimate by the estimator Fisher variance. We see the bias between the boosted and rest-frame  $f_{NL}$  measurements is significantly reduced and is now below the expected estimator variance (given in this case by the Fisher error).  \label{fig:fnl_withLinear}}
\end{figure}

Primordial non-Gaussianity searches aim to measure or constrain deviations from Gaussianity in the early universe. Such measurements are highly informative as, in many classes of early universe models, details of the primordial mechanisms are encoded into the deviations from Gaussianity. This subject has been extensively studied and we refer the reader to \citet{Chen_2010} for a detailed review of inflationary mechanisms and \citet{meerburg2019primordial} for a recent overview of the field. In this work we discuss the primordial bispectrum, which probes correlations between three Fourier modes of the primordial curvature perturbation, $\zeta(\mathbf{k})$. Assuming homogeneity and isotropy the primordial bispectrum, $B^\zeta(k_1,k_2,k_3)$ is defined as
\begin{align}
\langle \zeta(\mathbf{k}_1) \zeta(\mathbf{k}_2) \zeta(\mathbf{k}_3) \rangle = (2\pi)^3 \delta^{(3)}(\mathbf{k}_1+\mathbf{k}_2+\mathbf{k}_3)  \frac{3}{5} B^\zeta(k_1,k_2,k_3).
\end{align}
Due to the linearity of the early universe, measurements of the CMB bispectrum directly probe the primordial bispectrum.  Thus the primordial bispectrum is related to the reduced bispectrum by \citep{Komatsu_2001} 
\begin{align}
b^{X_1,X_2,X_3}_{\ell_1,\ell_2,\ell_3}  = \int r^2 \mathrm{d}r \prod \limits_i \int  \frac{2}{\pi}\mathrm{d}k_i g^{X_i}_T(k_i)j_{\ell_i}(k_i r)  B^\zeta(k_1,k_2,k_3),
 \end{align}
 where $g^{X_i}_T(k_i)$ are the transfer functions and $j_{\ell}(x) $ are the spherical Bessel functions. 
 
In this work we focus our discussion on the measurement of three primordial bispectra. Specifically we consider the local bispectrum
\begin{align}
 B^{\rm local}(k_1,k_2,k_3) =  & 2 f^{\rm local}_{\mathrm{NL}}A_s^2k_p^{8-2n_s} \left[ \frac{1}{k_1^{4-n_s}k_2^{4-n_s}}+ \right. \nonumber \\  &\left. \frac{1}{k_2^{4-n_s}k_3^{4-n_s}}+\frac{1}{k_1^{4-n_s}k_3^{4-n_s}} \right],
\end{align}
where $A_s$ is the amplitude of primordial fluctuations at the pivot scale $k_p$, $n_s$ is the spectral tilt. The local bispectrum is of particular interest as it is an informative probe of the field content of the early universe - \citet{Maldacena_2003,Creminelli_2004} has shown that the amplitude of the local bispectrum for all single field, slow roll inflationary models is slow roll suppressed. Thus measurement of a large amplitude of this type of non-Gaussianity would be highly informative.

The other two models we consider are the equilateral bispectrum,
\begin{align}
&B^{\rm{ equil}}(k_1,k_2,k_3) = 6 A_s^2 k_p^{8-2n_s} f^{\rm {equil}}_{\mathrm{NL}}\left(-\frac{1}{k_1^{4-n_s}k_2^{4-n_s}}  \right. \nonumber \\ &\left.  - \frac{1}{k_2^{4-n_s}k_3^{4-n_s}}-\frac{1}{k_1^{4-n_s}k_3^{4-n_s} }-\frac{2}{(k_1k_2k_3)^{2(4-n_s)/3} } \right. \nonumber \\ & \left.+\left[\frac{1}{k_1^{(4-n_s)/3}k_2^{(4-n_s)/3}k_3^{2(4-n_s)/3}} +\mathrm{5 perm.}\right] \right),
\end{align}
and  orthogonal bispectrum,
\begin{align}
&B^{\rm{ orth}}(k_1,k_2,k_3) = 6 A_s^2 k_p^{8-2n_s} f^{\rm {orth}}_{\mathrm{NL}}\left(-\frac{3}{k_1^{4-n_s}k_2^{4-n_s}} \right. \nonumber \\ &\left. - \frac{3}{k_2^{4-n_s}k_3^{4-n_s}} -\frac{3}{k_1^{4-n_s}k_3^{4-n_s} }-\frac{8}{(k_1k_2k_3)^{2(4-n_s)/3} } \right. \nonumber \\ & \left.+\left[ \frac{3}{k_1^{(4-n_s)/3}k_2^{(4-n_s)/3}k_3^{2(4-n_s)/3}} +\mathrm{5 perm.}\right] \right).
\end{align}
These types of non-Gaussianity are common in early universe models with strong non-linear dynamics \citep{Creminelli2006,Senatore2010}. These types of non-Gaussianity have been extensively searched for in CMB data sets with the current best constraints coming from the \textit{Planck} satellite \citep{planck2018_IX,Smith_2009,Bennett_2013}.

In this section we use a set of 160 temperature and E mode CMB maps generated using the \textit{Pixell} library. These maps are lensed CMB maps, which included the appropriate correlation between the lensing field and the integrate Sachs-Wolfe effect \citep{1968Rees,1967Sachs}. The maps are for differential thermodynamic measurements at 150 GHz and are computed as described in Section \ref{sec:boostingMethods}. We use $\ell_{\rm{ max}}=3000$ and consider the cosmic variance limited case (zero noise and instrumental effects). Considering the cosmic variance limited case allows us to assess the worse-case scenario for the impact of these effects. Note that we perform this analysis with an $\ell_{\rm{max}}=3000$ rather than higher as this is the range relevant for upcoming CMB experiments and it is highly computationally intensive.

We split the simulations into two sets: 80 simulations are used to estimate the linear term via an ensemble average, and 80 simulations are used as mock-observations. We consider a mask that only includes the region of sky within $60^\circ$ of the boost direction. Finally we also remove the induced dipole from our maps, as this is removed from CMB data analyses \citep{Planck2018III}.
First we consider applying bispectrum estimators without the linear term. Whilst this exercise is purely academic, as in the analysis of experimental data a linear term is almost always subtracted, we find that there is an interesting effect present in the analysis without a linear term. The results are shown in Figure \ref{fig:fnl_noLiinear}. We see that the orthogonal and local bispectra, all bispectra with a large squeezed contribution, exhibit a systematic bias. We have normalized these templates by the Fisher variance to set a similar scale for the different measurements. However the measured variance is significantly larger, due to the exclusion of the linear term, such that these biases are unobservable.
\citet{Catena_2013b} found no bias between their simulated boosted and rest-frame maps. We identify the source of this discrepancy as arising from the difference in measurement types - differential thermodynamic measurements compared to temperature measurements. When repeating our computation for temperature measurements we find that this bias disappears. We return to the origin of this bias after considering estimators that include the linear term. 

In Figure \ref{fig:fnl_withLinear} we repeat our analysis including the linear term, which we compute using 80 held-out simulations. We see that the biases are all significantly reduced and are now at the level of $0.2 \sigma$. This large reduction in bias elucidates the origin of the bias seen in the measurements without the linear term. When computing the boost for differential thermodynamic measurements, in addition to the dipole contribution there is a Doppler induced quadrupole. This quadrupole combines with the off-diagonal power spectrum contributions arising from the masking (and in experimental data inhomegenous noise) to produce a bispectrum contribution as 
\begin{align}\label{eq:quadBias}
\langle a_{\ell_1 m_1} a_{\ell_2 m_2} a_{\ell_3 m_3} \rangle= \langle a_{\ell_1 m_1} \rangle \langle a_{\ell_2 m_2} a_{\ell_3 m_3} \rangle = \langle a_{\ell_1 m_1} \rangle C_{\ell_2m_2, \ell_3 m_3}.
\end{align}
When we include the linear term in our analysis we remove most of the the off-diagonal power spectrum contributions and thereby remove the bias. Note that we still see a small residual bias, this  arises as our linear term does not provide a complete cancellation of the off-diagonal power spectrum, likely due to that fact we could only use  80 simulations to compute this term - due to the high computational cost.

Next we consider whether the boost introduces extra variance into the bispectrum estimator, similar to how lensing introduces extra variance \citep{Coulton_2020,Babich_2004}. We are motivate to explore this question from the results seen in \citet{Catena_2013} who find  a scatter between the boosted and rest-frame $f_{NL}$ measurements at the level of 30$\%$ of their estimator variance.  In Figure \ref{fig:fnl_withLinear} we see a similar level of scatter. However our analysis includes polarization data, whilst  \cite{Catena_2013} used only temperature, smaller scales and we include the linear term in our analysis, hence have a more optimal estimator.
 These three factors mean that whilst we also see $30\%$ scatter our error bars are a factor of two smaller. 
 That means that as the constraining power of estimator improves, the scatter between the boosted and rest-frame maps decreases. This implies that boosting does not cause additional noise in the bispectrum estimator, rather the scatter likely arises as the boost results in a decorrelation between the boosted-frame and rest-frame maps, as is seen by computing the cross correlation between the boosted-frame and rest-frame power spectra. This is similar to the effect seen in the one-point function analyses.

\begin{figure}
    \centering
    \includegraphics[width=\linewidth]{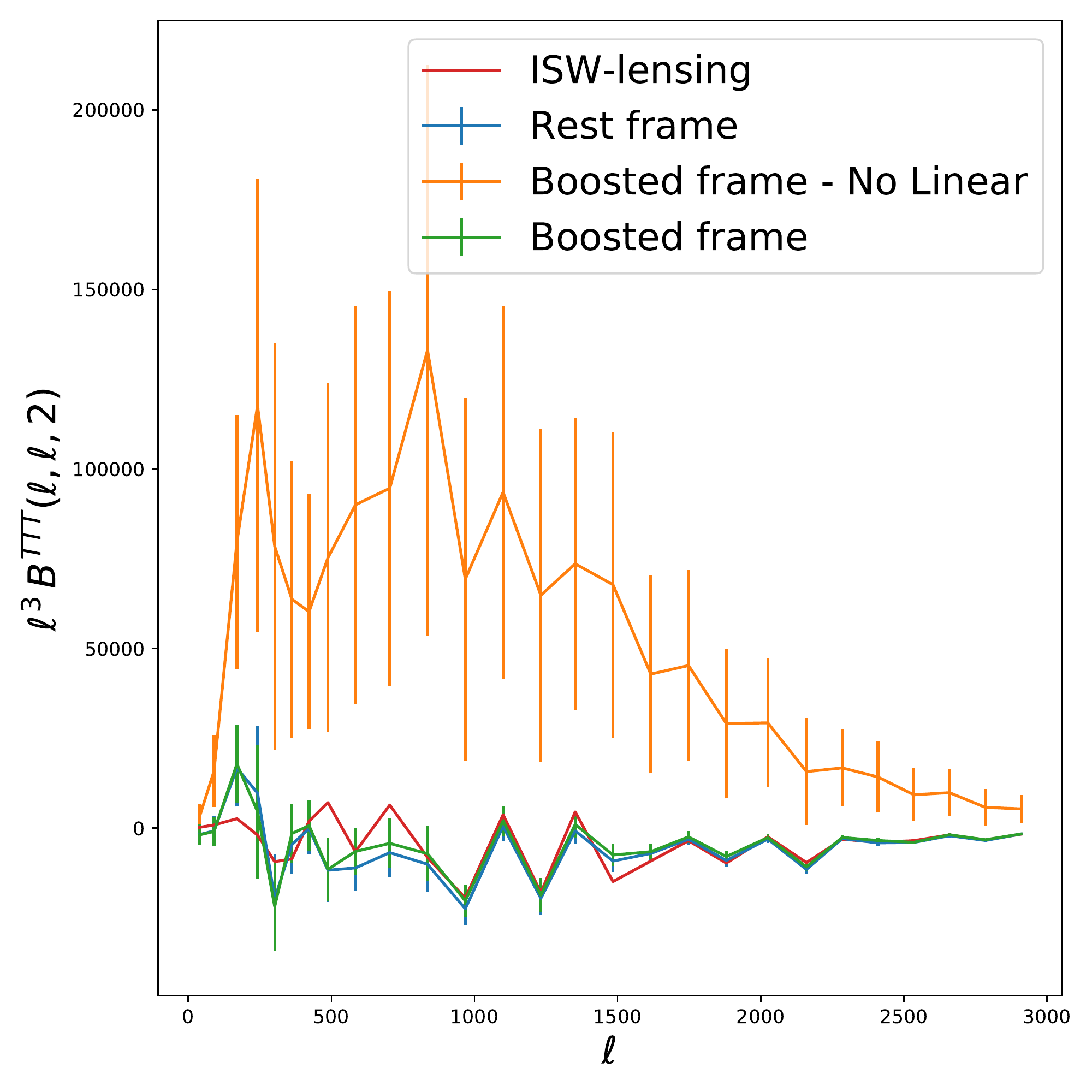}
    \caption{A slice of the squeezed ISW-lensing bispectrum for the boosted and rest-frame CMB maps. We plot results both included and excluding the linear term. Without the linear term we see the leakage of the quadrupole. With the linear term, we see that we recover the theoretical ISW-lensing bispectrum. We note that the ISW-lensing bispectrum is a small signal hence the large scatter even after averaging over 80 simulations. \label{fig:squeezedBisp}}
\end{figure}
\subsection{Integrated Sachs-Wolfe -lensing bispectrum}\label{sec:ISWlensingBispectrum}
The correlation between the ISW and lensing effects results in a non-zero bispectrum in the CMB that is known as the ISW-lensing bispectrum \citep{Goldberg_1999}.  This bispectrum has been measured by the \textit{Planck} satellite at {$\sim 3\sigma$} \citep{Planck2015XXI}. This bispectrum is an interesting probe of dark energy, spatial curvature and modified gravity \citep{Crittenden1996,Hu_2001,Kamionkowski_1996}. Further it is interesting to examine here as its has a shape that is similar to the local primordial bispectrum \citep{Lewis_2011}  and is computationally cheap to simulate. These two facts mean that it is interesting to study how boosting impacts measurements of the ISW-lensing bispectrum as the primordial bispectra will likely be effected in a similar manner. We included this source of non-Gaussianity in the simulations used in the previous section.

As for the primordial bispectra we measure the amplitude of ISW-lensing bispectrum both with and without the linear term. The results seen in Figures \ref{fig:fnl_noLiinear} and \ref{fig:fnl_withLinear} show the same pattern as the primordial local-type non-Gaussianity. This is expected as the bispectra templates are similar.
In absolute terms, when including the linear term, we find amplitudes of the ISW-lensing bispectra as $\hat{f}_\text{NL}= 0.98 \pm 0.03$, from the average of the rest-frame maps and  $\hat{f}_\text{NL}= 1.00 \pm 0.03$ for the boosted maps. In Figure \ref{fig:squeezedBisp} we plot a squeezed slice of the boosted and rest-frame bispectra as well as the theory expectation. For the estimates without the linear term, we see a systematic offset between the boosted and rest-frame bispectra. This is the result of the quadurpole effect discussed in the previous section. This offset  is removed when we include the linear term.  For all other configurations we see no statistically significant difference, in line with our explanation. We can be further assured that there is no systematic offset as the $f_{NL}$ results, including the linear term, are consistent and this is a more stringent test than a by-eye comparison.

The results of this section indicate that there is no observable impact of the boost on bispectrum measurements up to $\ell_{\rm{max}}=3000$, the relevant range for proposed experiments. Given our results, and the physics of the boost, we expect there to be no impact on higher resolution measurements as well.

\subsection{Sunyaev-Zeldovich effects}
\begin{figure}
    \centering
    \includegraphics[width=\linewidth]{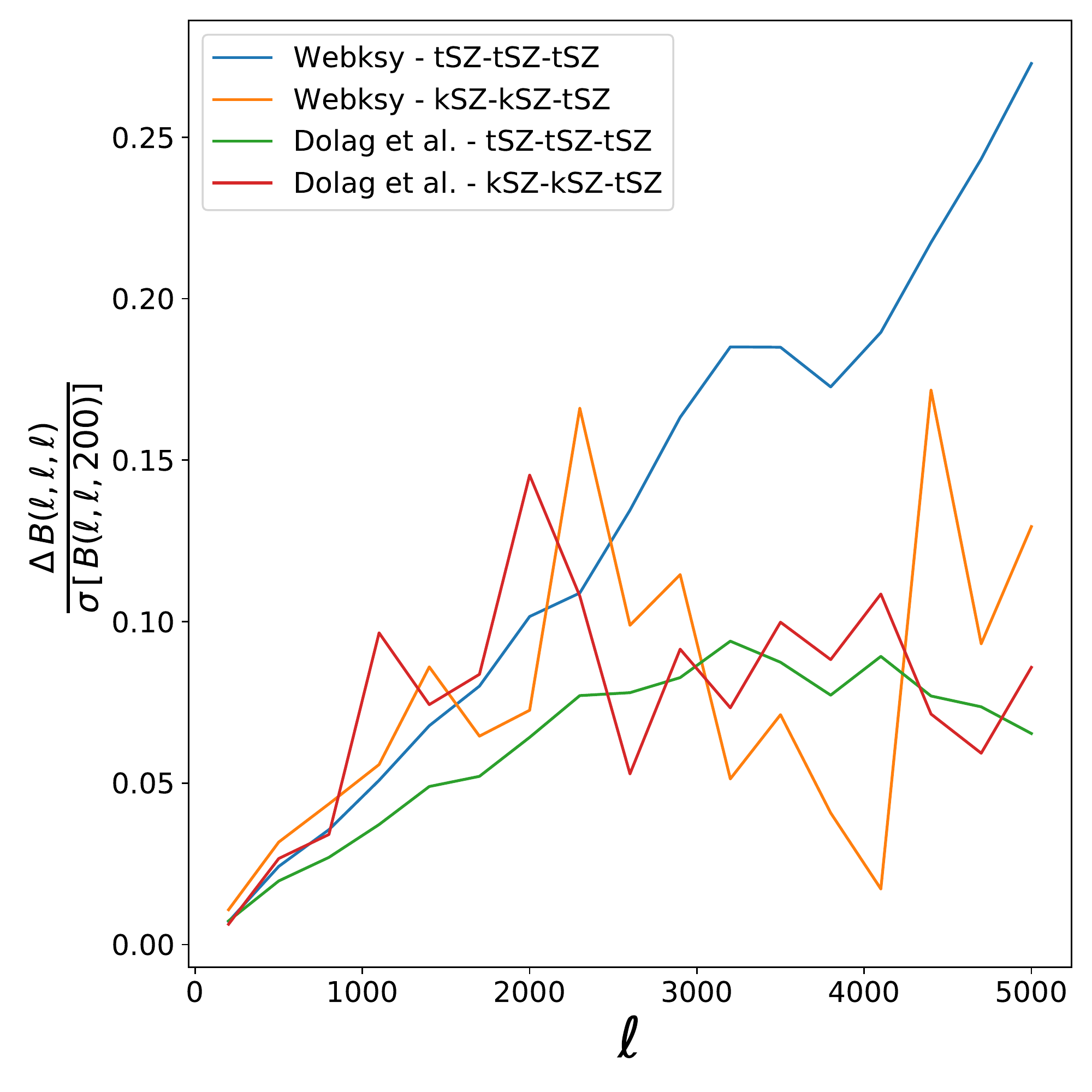}
    \caption{A plot showing the significance of the difference between the boosted-frame and rest-frame tSZ and kSZ bispectra for the equilateral configuration. We compute the mean difference between our 12 boosted patches and compare this with the variance of the rest-frame bispectrum. \label{fig:bispectra_sig}}
\end{figure}

In this section we explore the impact of Doppler boosting and aberration effects on the thermal and kinetic Sunyaev-Zeldovich bispectra. Our experimental setup is the same as in Section \ref{sec:onePointAn}:  we divide the sky into twelve healpix pixels, boost each one in the direction of it centre and consider observations at 143 GHz. This allows us to estimate the worse case impact of the boosting effects.

In Figure \ref{fig:bispectra_sig} we plot the significance of the difference between the rest-frame and boosted-frame maps for the tSZ-tSZ-tSZ and kSZ-kSZ-tSZ bispectra. These are the only non-zero bispectra between these two fields (all configurations involving an odd number of kSZ maps will be zero when averaged over realizations). We see that the amplitude of these bispectra is altered by the boost, however the effect is significantly below cosmic variance on all scales. We did not add the primary CMB dipole to the kSZ map and the kSZ and tSZ induced monopoles are very small  \citep{Fixsen_1996,Hill_2015}; combined this means we do not see the same quadrupole leakage effects that are seen in the CMB bispectrum plots.

We concluded that the impact of doppler boosting on bispectrum measurements of the tSZ and kSZ effects can be safely neglected.

\section{4-POINT STATISTICS} \label{sec:fourPointAn}
Whilst computing the full four-point function, or its harmonic equivalent, is computational prohibitive, statistics that target specific subsets have been extensively studied. The two most explored statistics are those to probe primordial trisipectra, for example $g_{\mathrm{NL}}$ and $\tau_{\mathrm{NL}}$ \citep{smith2015optimal,fergusson2010optimal,Kogo_2006,Komatsu2002}, and statistics used to probe the lensing induced trispectrum  \citep{Seljak_1996,LEWIS_2006}. In this work we focus on the later set of statistics. 
\subsection{Gravitational Lensing Trispectrum} \label{sec:lensing4point}
We start by transforming the deflection equation, Eq. \ref{eq:aberationLocations}, to harmonic space and Taylor expanding it. We see that, in addition to modifying the power spectrum of the observed anisotropies, lensing couples different spherical harmonic coefficients as
\begin{align} \label{eq:lensingCoupling}
&a(\boldell) \approx \bar{a}(\boldell) \non  &-\int\frac{\mathrm{d}^2\ell_1}{(2\pi)^2} \frac{\mathrm{d}^2\ell_2}{(2\pi)^2} \boldell_1\cdot \boldell_2 \bar{a}(\boldell_1)\phi(\boldell_2) (2\pi)^2\delta^{(2)}(\mathbf{L}-\boldell_1-{\boldell_2}).
\end{align}
Note that we work in the flat-sky regime as this will be helpful for building intuition later and we remind the reader that the bar denote unlensed quantities.  It is clear that this means the anisotropies have a non-trivial four-point function and are hence non-Gaussian. The lensing induced non-Gaussianity is a rich source of cosmological information as its properties are determined by the lensing potential, which depends on the line-of-sight integrated matter in the universe. 

Using the off-diagonal coupling, Eq. \eqref{eq:lensingCoupling}, we can derive a quadratic estimator, for the lensing potential,$\hat{\phi}$, as \citep{Okamoto_2003,Hu_2002}
\begin{align}\label{eq:quadraticEst}
\hat{\phi}(\mathbf{L}) =& \mathcal{R}_L \int\frac{\mathrm{d}^2\ell}{(2\pi)^2} \frac{ a(\boldell)a(L -\boldell) \left[C_{\ell}\boldell\cdot\mathbf{L}+C_{|\mathbf{L}-\boldell|}\mathbf{L}\cdot(\mathbf{L}-\boldell) \right]}{2(C^{\mathrm{Tot}}_\ell)(C^{\mathrm{Tot}}_{|\mathbf{L}-\boldell|})},
\end{align}
where $C^{\mathrm{Tot}}$ is the total observed power spectrum, including instrument noise, and $ \mathcal{R}_L $ is a normalisation constant that can be analytically approximated by
\begin{align}\label{eq:lensingNorm}
     \mathcal{R}_L = L^2 \left [\int\frac{\mathrm{d}^2\ell}{(2\pi)^2} \frac{ \left[C_{\ell}\boldell\cdot\mathbf{L}+C_{|\mathbf{L}-\boldell|}\mathbf{L}\cdot(\mathbf{L}-\boldell) \right]^2}{2(C^{\mathrm{Tot}}_\ell)(C^{\mathrm{Tot}}_{|\mathbf{L}-\boldell|})} \right]^{-1}
\end{align}

 We follow \citet{Hanson_2011} and use lensed power spectra in the numerator of the estimator as this suppresses higher order biases. We can then extract cosmological information from this reconstructed field by studying its moments, primarily the power spectrum (which is therefore an implicit four-point function).

Existing experiments have already made detections of the lensing power spectrum  \citep{Sherwin_2017,Planck2018VIII,Bianchini_2020,Fa_ndez_2020} and upcoming and on-going experiments plan to use precise measurements of the lensing potential to constrain cosmological parameters, such as the neutrino mass \citep{Mishra_Sharma_2018,Ade_2019,abazajian2016cmbs4}. Given the high level of expected precision, it is important to examine whether Doppler effects will impact and bias lensing measurements.

It is immediately evident from a comparison of Eq. \eqref{eq:aberationLocations} and Eq. \eqref{eq:lensingMapLevel} that lensing and aberration effects produce identical effects, remapping of the observed anisotropies. A fundamental difference between the two is that the aberration field is a dipolar field whereas the lensing field has power on all scales. This similarity means that quadratic estimators, like Eq. \eqref{eq:quadraticEst}, can be used to measure the aberration effect as was done in \citet{Aghanim:2013suk}. Further it means that a measurement of the lensing dipole will be biased. In this work we wish to explore whether any additional biases are produced.  

We use a lensing pipeline similar to one used in analysis of ACT data \citep{Sherwin_2017}. Our pipeline has three stages. First we apply our quadratic estimator to the simulated map. Note that the total power spectrum used in the denominator of our estimator is different for the boosted-frame and rest-frame reconstructions, as the Doppler and aberration effects alter the total level of power in the maps. Next we subtract a component, known as the the mean field, for our quadratic estimator. Thus we have
\begin{align}
\bar{\hat{\phi}}(\mathbf{L}) = \hat{\phi}(\mathbf{L}) -\langle \hat{\phi}(\mathbf{L})  \rangle.
\end{align}
where $\langle \hat{\phi}(\mathbf{L})  \rangle$ is the mean field term and is simply computed by computing the average value of the estimator. The mean field accounts for the fact that masking, along with inhomogeneous noise in dataset, introduces a non-zero expectation value of the quadratic lensing estimator. This will bias cosmological inferences if not accounted for. As this term depends on the power in the map it could differ between the boosted and rest-frame maps. Whilst we use separate mean-fields for the boosted and rest-frame analyses we tested and found no differences in our results if we used the same mean-field, the rest-frame mean-field, in both analyses. 

Next we compute the power-spectrum of the mean-field subtracted field. This gives a biased estimate of the lensing potential. The largest bias, known as the $N^{(0)}$ bias, arises as the power spectrum is actually a four-point function that has Gaussian contributions even in the absence of gravitational lensing.  We compute this bias and remove it using the method developed in \citet{Namikawa_2013}. The leads to
\begin{align}
C^{{\phi}{\phi}}(\mathbf{L}) =C^{\bar{\phi}\bar{\phi}}(\mathbf{L}) - N^{(0)}(\mathbf{L}).
\end{align}
This estimate is still a biased estimate of the true lensing potential power spectrum. The next most significant bias, known as the $N^{(1)}$ bias \citep{Story_2015,Hanson_2011}, depends linearly on the power spectrum of the lensing potential. In data analyses this bias is also subtracted, however, unlike the previous biases, residuals from this subtraction must also be accounted for in the likelihood. This occurs as the N$^{(1)}$ bias depends on the lensing power spectrum and so differences between the cosmology used to subtract the bias and the true cosmology will lead to biases in inferred cosmological parameters. Given this complication, and the significant computational overhead required to compute this bias, we do not remove it but will comment on the expect impact of Doppler and aberration effects below.

We first apply the lensing pipeline to the rest-frame map and then repeat the procedure on the boosted map. To study these biases we use the same experimental setup as in Section \ref{sec:threePointAn}: noise-free, differential thermodynamic measurements at 150 GHz with a mask that retains all the sky within $60^\circ$ of the boost direction. We again use an $\ell_{\rm {max}}=3000$ as is the range of interest for upcoming CMB experiments. We only use temperature maps in this analyses and defer an analysis including polarization to future work.

\begin{figure}
    \centering
    \includegraphics[width=\linewidth]{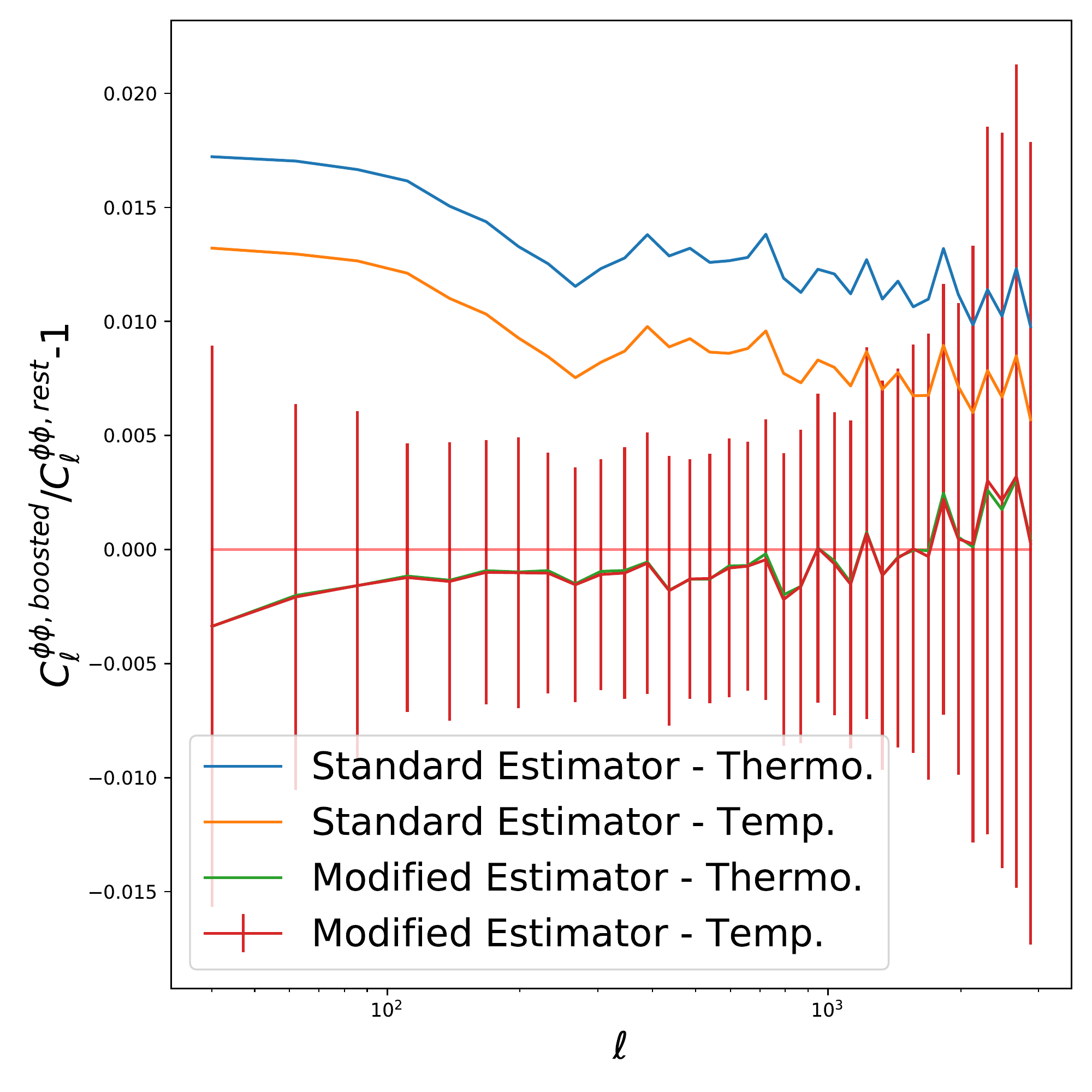}
    \caption{The fractional difference between the lensing power spectrum measured a set of boosted and frame simulations. We use a mask that retains only the portion of the sky within $60^\circ$ of the boost direction. We average our measurements over 140 noise-free CMB simulations. In orange, we see that using the standard lensing estimator to estimate the lensing potential results in a biased measurement of the lensing power spectrum. We propose a modified estimator, see the discussion surrounding Eq.\eqref{eq:modifedBispectrum}, and plot the results of that estimator in blue. We see this provides an unbiased measurement of the lensing potential power spectrum. \label{fig:cl_kk}}
\end{figure}

In Figure \ref{fig:cl_kk} we plot the fractional difference between the boosted and rest-frame lensing power spectrum. We see that the boosted power spectrum is biased and is $\sim1.5\%$ larger than the true lensing power-spectrum at all scales. Note that this bias is weaker for the temperature measurements (compared to the differential thermodynamic measurements shown here) as the modulation effect is stronger in the differential thermodynamic measurements. 

There is a simple heuristic for this bias: at small scales the effect of the Doppler and aberration effects is primarily to increase the small scale power in the direction of the boost. The quadratic estimator, Eq. \eqref{eq:quadraticEst}, weighs modes by the rest-frame CMB power spectrum. This is incorrect as boosting has altered the power. If unaccounted for, this difference will induce a multiplicative bias in the measurements, proportional to the fraction of increased power. 

More formally this effect can be thought of as bias akin to the bias that occurs if unlensed power spectra are used in the quadratic estimator,  Eq. \eqref{eq:quadraticEst}, \citep[see][ for more details]{Hanson_2011}. To understand this effect, first we examine the structure of the quadratic estimator. The general structure of the quadratic estimator is \citep{darwish2020density}
\begin{align}
 & \hat{\phi}(\mathbf{L}) = \mathcal{R}_L \int\frac{\mathrm{d}^2\ell_1}{(2\pi)^2} \frac{\mathrm{d}^2\ell_2}{(2\pi)^2} \frac{ a(\boldell_1)a(\boldell_2) (2\pi)^2\delta^{(2)}(\mathbf{L}+\boldell_1+{\boldell_2})}{2C^{\mathrm{Tot}}_\ell C^{\mathrm{Tot}}_{|\mathbf{L}-\boldell|}} \non & \times B^{\phi TT}(L,\boldell_1,\boldell_2),
\end{align}
where $B^{\phi TT}(L,\ell_1,\ell_2)$ is the bispectrum between the temperature fields. For the rest-frame sky, this bispectrum is, in the flat-sky regime,
\begin{align}
&\langle \phi(\boldell_1) a(\boldell_2)a(\boldell_3)\rangle =   (2\pi)^2\delta^{(2)}({\boldell_1}+{\boldell_2}+\boldell_3)B^{\phi TT}(\ell_1,\ell_2,\boldell_3).\nonumber \\
									             = &  -(2\pi)^2\delta^{(2)}({\boldell_1}+{\boldell_2}+\boldell_3)C^{\phi\phi}_{\ell_1}C^{TT}_{\ell_2}\boldell_1\cdot \boldell_2 +\ell_2\xleftrightarrow{} \ell_3.
\end{align}
In the presence of the Doppler, aberration and lensing effects the observed anisotropies are
\begin{equation}\label{eq:lensingAndAbMapLevel}
 \tilde{T}(\mathbf{n}) =M(\mathbf{n}) \bar{T}(\mathbf{n}+\mathbf{s}+\nabla \phi(\mathbf{n})),
\end{equation}
where $\mathbf{s}$ is the remapping caused by the aberration effects and $M(\mathbf{n})$ is the modulation. We see immediately that the effects of lensing and aberrations commute, whilst the modulation is more complicated. Transforming to harmonic space we can recompute the bispectrum to obtain
\begin{align}
\langle \phi(\boldell_1) \tilde{a}(\boldell_2)\tilde{a}(\boldell_3)\rangle = & -C^{\phi\phi}_{\ell_1}\int \frac{\mathrm{d}^2\ell_A}{(2\pi)^2} \frac{\mathrm{d}^2\ell_B}{(2\pi)^2}\tilde{M}(\boldell_2+\boldell_A) \non &  \times \tilde{M}(\boldell_3+\boldell_1-\boldell_A) \bar{C}^{TT}_{\ell_A}\boldell_1\cdot \boldell_A +\ell_2\xleftrightarrow{} \ell_3,
\end{align}
where $\tilde{M}(\boldell)$ is the harmonic transform of the modulation effect and $\bar{C}$ is the temperature power spectrum including the aberration effect. Now, by noting that the modulation kernel only couples nearby scales and hence we can approximate $\tilde{M}(\boldell) \sim 0$ for $\ell \gtrsim 4$.  With this approximation the bispectrum has the following approximate form
\begin{align}\label{eq:modifedBispectrum}
\langle \phi(\boldell_1) \tilde{a}(\boldell_2)\tilde{a}(\boldell_3)\rangle = 
 & -C^{\phi\phi}_{\ell_1}\boldell_1\cdot \boldell_2 \tilde{C}^{TT}_{\ell_2} +\ell_2\xleftrightarrow{} \ell_3 ,
\end{align}
where $ \tilde{C}^{TT}$ is the CMB power spectrum including the Doppler and aberration effects. Given the structure of the bispectrum we can immediately see that the quadratic estimator used above is biased as it uses the incorrect bispectrum. Note that the bias actually arises in the estimator normalization, Eq. \ref{eq:lensingNorm}. We normalize the estimator by the estimator weights squared - one factor removes the weights used in the estimator and the second to remove the $C_\ell \boldell_1\cdot \boldell_2$ factor from the signal bispectrum. The bias thus arises as the true bispectrum is different from the assumed one and so produces an incorrect normalization - this explains why the bias is approximately independent of scale. This analysis allows more insight to the bias, lensing estimators really measure the product of  $\phi_\ell$ and $C_\ell$ and the lensing potential is obtained by dividing by  $C_\ell$. Thus if the CMB power spectrum on the patch of observation is misestimated, the inferred lensing potential will be biased.

We can simply correct this by using the boosted and lensed power spectrum in the numerator of the quadratic estimator. In Figure \ref{fig:cl_kk} we plot the lensing power spectrum estimated with this modified estimator. We see that the new estimator is unbiased and gives the same estimator as the estimator applied to rest-frame simulations. We find that this is the case for both the temperature and thermodynamic simulations.  We note that this result implies that the Doppler and aberration effects impact the N$^{(1)}$ bias in an identical manner and so these should be computed using boosted power spectra. 

\section{Conclusions}\label{sec:conclusions}

Given the high precision of existing and upcoming CMB experiments, it is necessary to ensure all physical processes and systematic effects are understood to a similar level. In this work we have examined the impact of our motion on statistical probes of the CMB and SZ galaxy clusters' maps.
In order to reach this goal, we utilized tSZ and kSZ maps from two different types of simulations: 
semianalytical and fully hydrodynamical.
First, we scrutinized current strategies for boosting simulated maps, and then we computed the impact of our motion on a variety of statistical probes.

\vspace{0.1cm}
For the first part our results can summarized as follows:

$\bullet$ We computed to what extent boosting corrections may be important in interpreting a variety of upcoming experiments. Table \ref{tab:survey_aberration} shows that, even for large area coverage experiments,  these can be significant when considering patches. For example,  Simons Observatory and CMB-S4 will have similar boosting factors over each of the two patches of sky observable from the southern hemisphere (the available sky is divided into two by the Galaxy) and thus accounting for these effects is important for internal consistency tests.

$\bullet$ We validated boosting codes by 
comparing the temperature and polarization power spectra of boosted maps obtained with two different boosting codes (\textit{Cosmoboost}  and \textit{Pixell}).
We showed good agreement between outcomes up to $\ell \simeq 5000$.
In terms of computational time, the resources for boosting a single map are approximately similar between the two codes, with \textit{Pixell} being faster at a larger memory footprint. For boosting multiple maps \textit{Cosmoboost} was found to be a factor of a few faster.

$\bullet$ We derived  an analytical formula to compute the  boosted spectrum  for the map of a signal with  a generic frequency dependence  (see Eq. \ref{eq:jeongExtended}). This is a generalization of the formula 
introduced by \citet{Jeong:2013sxy} relative to CMB temperature.
This generalization  is important not only for modelling the tSZ power spectrum, but also for the primary CMB observations, which tend to be measured by  differential intensity measurements that are frequency dependent.
This new formula shows a good agreement with the results of $Cosmoboost$ for SZ maps and both codes for CMB maps.
Including the frequency dependence is crucial for modelling the power spectra of these observables:
for the CMB power spectrum, the frequency dependence can significantly increase the size of the modulation term, increasing the correction form the boost by a factor of $\sim30\%$. Thus including these effects is necessary to avoid biased inferences. 

$\bullet$ In some occasions, such as when examining the tSZ signal near its null frequency, higher order derivatives of the frequency response can be important. We provided and validated an extended power spectrum boosting formula, Eq. \ref{eq:jeongExtendedWith2ndFreq}. This formula includes the second order frequency derivative and was found to provide a reasonably accurate match to \textit{Cosmoboost} over all frequencies considered.

\vspace{0.1cm}

The second part of our work focused on studying how the Doppler and aberration effects impact
the statistics of both CMB and SZ effects.
The case of the CMB power spectrum has extensively been discussed in the literature, so we specifically focused on the SZ  power spectrum  and the  non-Gaussian statistics  of both CMB and SZ effects. 

We found the following results:

$\bullet$ For both tSZ and kSZ power spectra, the average expected effect of the boost is small (0.5-1\% in the direction of the observer's velocity, considering 1/12 of the sky at 143 GHz), and subdominant to the variance amongst the  patches in the unboosted frame. However it can be an O(1) correction to the signal when near the tSZ null. The precise importance of these effects near the null will depend upon the experimental bandpass, as well as the specific area considered in the data analysis. Results from different simulations are consistent in showing the effect.

$\bullet$ The boosting does not produce an appreciable change in the CMB skewness and kurtosis or its variance. Further there is no evident skewness or kurtosis signature  induced by the boosting on the tSZ and kSZ $a_{\ell m}$ distributions, other than possibly a small change in the tSZ kurtosis for Dolag's simulations.

$\bullet$ We considered the impact on the CMB, kSZ and tSZ bispectra and found that it is unimportant. There is a novel, but unobservable impact on bispectrum estimators; we found that the modulation induced quadurople can leak into measurements of the CMB bispectrum, producing a systematic bias. Whilst this effect is unobservably small, we found that the bias is automatically removed by in inclusion of the bispectrum 'linear' term, which is already included in most bispectra analyses as a method to reduce the estimator's variance.

$\bullet$ Estimators of the lensing power spectrum are found to be impacted by a multiplicative bias from the boosting effect. This arises as lensing estimators measure the product of $\phi_\ell$ and  $C_\ell$ and the Doppler and aberration effects modify the power spectrum of the CMB anisotropies. Thus estimators that use the rest frame $C_\ell$, rather than the boosted spectra on the observed patch, will infer a biased value $\phi_\ell$. We present a simple formula to correct this bias that can be implemented at minimal extra computational cost and reduces the bias to a negligible level.

We note that the results of the lensing and bispectrum analyses used a mask that contained only the quarter of the sky in the direction of the boost. If a patch of the sky that is less well align with the boost is observed the effects discussed will be suppressed. However, as is seen in Table \ref{tab:survey_aberration}, experiments such as the Simons Observatory and CMB-S4 will have similar boosting factors over each of the two patches of sky observable from the southern hemisphere (the available sky is divided into two by the Galaxy) and thus accounting for these effects is important for internal consistency tests.

\section{Acknowledgements}
The authors thank Daan Meerburg, Jens Chluba, Adriaan J. Duivenvoorden, Marcelo Alvarez and Anthony Challinor for fruitful discussions. W.R.C. acknowledges support from the UK Science and Technology Facilities Council (grant number ST/N000927/1), the World Premier International Re- search Center Initiative (WPI), MEXT, Japan and the Center for Computational Astrophysics of the Flatiron Institute, New York City. The Flatiron Institute is supported by the Simons Foundation.
EP and SY were  supported by NSF grant AST-1910678 and NASA award 80NSSC18K0403. EP is also a Simons Foundation Fellow.
The numerical simulations were partially generated at the USC high performance computing facility (CARC).
EP and WC thanks the Aspen Center for Physics, which is supported by National Science Foundation grant PHY-1607611, for hospitality during the preparation of this work.
SF acknowledges  the support of the USC Provost fellowship during the preparation of this work.

KD acknowledges support by the COMPLEX project from the European Research Council (ERC) under
the European Union’s Horizon 2020 research and innovation program grant agreement ERC-2019-AdG 882679 as well as support by the Deutsche Forschungsgemeinschaft (DFG, German Research Foundation) under Germany’s Excellence Strategy - EXC-2094 - 390783311. The calculations for the hydro-dynamical simulations were carried out at the Leibniz Supercomputer Center (LRZ) under the project pr86re. We are especially grateful for the support by M. Petkova through the Computational Center for Particle and Astrophysics (C2PAP) and the support by N. Hammer
at LRZ when carrying out the Box0 simulation within the Extreme Scale-Out Phase on the new SuperMUC Haswell extension system.

\bibliographystyle{mnras}
\bibliography{cosmobib.bib}

\end{document}